\documentclass{aa}
\usepackage{graphicx}

\begin{document}

\def\lesssim{{_ <\atop{^\sim}}}
\def\grtsim{{_ >\atop{^\sim}}}
\def\fd{\mbox{$f_{\rm d}$}}
\def\fg{\mbox{$f_{\rm g}$}}
\def\kms{\mbox{kms$^{-1}$}}
\def\vdvt{\mbox{$(V_{\rm d,m}/V_{\rm t,m})$}}
\def\mdynmd{\mbox{$M_{\rm dyn}/M_{\rm bar}$}}
\def\mdynms{\mbox{$M_{\rm dyn}/M_{\rm s}$}}
\def\mdynl{\mbox{$M_{\rm dyn}/L_{B}$}}
\def\mdynlK{\mbox{$M_{\rm dyn}/L_{ K}$}}
\def\mdyn{\mbox{$M_{\rm dyn}$}}
\def\hd{\mbox{$h_{\rm d}$}}
\def\hB{\mbox{$h_{B}$}}
\def\hK{\mbox{$h_{K}$}}
\def\hs{\mbox{$h_{\rm s}$}}
\def\hl{\mbox{$h_{\lambda}$}}
\def\rma{\mbox{$r_{\rm peak}$}}
\def\mv{\mbox{$M_{\rm v}$}}
\def\md{\mbox{$M_{\rm bar}$}}
\def\mg{\mbox{$M_{\rm g}$}}
\def\ms{\mbox{$M_{\rm s}$}}
\def\mh{\mbox{$M_{\rm dm}$}}
\def\msun{\mbox{M$_\odot$}}
\def\sigd{\mbox{$\Sigma_{\rm d,0}$}}
\def\sigl{\mbox{$\Sigma_{\rm \lambda,0}$}}
\def\sigB{\mbox{$\Sigma_{B,0}$}}
\def\sigK{\mbox{$\Sigma_{K,0}$}}
\def\sigs{\mbox{$\Sigma_{\rm s,0}$}}
\def\mul{\mbox{$\mu_{\rm \lambda,0}$}}
\def\muK{\mbox{$\mu_{K,0}$}}
\def\vd{\mbox{$V_{\rm d,m}$}}
\def\vt{\mbox{$V_{\rm t,m}$}}
\def\vtsq{\mbox{$V^{2}_{\rm t,m}$}}
\def\W{\mbox{$W_{20}$}}
\def\LCDM{\mbox{$\Lambda$CDM}}

\title{The luminous and dark matter content of disk galaxies}

\author{Jes\'us Zavala\inst{1}, Vladimir Avila-Reese\inst{1}, H\'ector Hern\'andez-Toledo\inst{1} \and Claudio Firmani\inst{2}}

\offprints{V. Avila-Reese}
 
\institute{Instituto de Astronom\'\i a--UNAM, A.P. 70-264, 04510 
M\'exico, D. F., M\'exico \\
\email{jzavala,avila,hector@astroscu.unam.mx}
\and Osservatorio Astronomico di Brera, via E.Bianchi 46, I-23807 Merate, Italy \\
\email{firmani@merate.mi.astro.it}}

\titlerunning{Luminous and dark matter in disk galaxies}
\authorrunning{Zavala et al.}

\date{Received .............../ Accepted ...............}

\abstract{We have compiled a sample of disk galaxies with
available photometry in the $B$ and $K$ bands, velocity line-widths and HI
integral fluxes. Several parameters which trace the luminous, baryonic and 
dark matter contents were inferred. We investigated how these parameters do vary 
with different galaxy properties, and confronted the results with predictions 
of galaxy evolutionary models in the
context of the $\Lambda$ Cold Dark Matter (\LCDM) cosmogony. The
ratio of disk-to-total maximum circular velocity, \vdvt, depends mainly
on the central disk surface density \sigd\ (or surface brightness, SB),
increasing roughly as $\sigd^{0.15}$. While a fraction of high SB galaxies 
have a \vdvt\ ratio corresponding to the maximum disk solution, the low 
SB are completely dark matter dominated. The trend is similar for the models, 
although they have
slightly smaller \vdvt\ ratios than observations, in particular at
the highest SBs and when small baryon fractions are used.  
The scatter in the $\vdvt-\sigd$ plot is large. An analysis
of residuals shows that \vdvt\ tends to decrease as the galaxy is redder, more
luminous (massive), and  of earlier type. The models allow us to explain the 
physics of these results, which imply a connexion between halo structure and
luminous properties. The dynamical-to-baryon mass and dynamical mass-to-light 
($B$ and $K$) ratios at a given radius were also estimated.  All these ratios, 
for observations and models, decrease with \sigd\ (or SB) and {\it do not correlate} 
significantly with the galaxy scale, contrary to what has been reported in 
previous works, based on the analysis of rotation curve shapes. We discuss this 
difference and state the importance to solve the controversy on whether the dark 
and luminous contents in disk galaxies depend on SB or luminosity. 
The broad agreement between the models and observations presented here
regarding the trends of the dynamical-to-baryon matter
and mass-to-light ratios with several galaxy properties favors the
\LCDM\ scenario. However, the excess of dark matter inside the optical
region of disk galaxies remains as the main difficulty.
}

\maketitle

\section{Introduction}

The information about the distribution of luminous and dark matter in disk
galaxies, as well as on the correlations among the main parameters that
characterize both components, offers a fundamental clue to understand how
galaxies did form and evolve and what role did dark matter play in these
processes. Current models of disk galaxy formation and
evolution, based on the hierarchical Cold Dark Matter (CDM) scenario, make
certain predictions about these distributions and correlations
(e.g., Firmani, Avila-Reese \& Hern\'andez 1997; Dalcanton et al. 1997; Mo, Mao
\& White 1998; Avila-Reese, Firmani \& Hern\'andez 1998; van den Bosch 1998,
2000; Firmani \& Avila-Reese 2000; Avila-Reese \& Firmani 2000). These
predictions can be used to confront and interpret the available observations. 
This is the aim of the present paper, where we will compare some
global properties and correlations of observed galaxies with the model ones. 
From the point of view of the models, the galaxy properties are the result
of the combination of some fundamental (cosmological) parameters, which follow 
continuous statistical distributions. Therefore, an observational sample
as complete as possible in luminosities, surface brightnesses, morphological
types and integral colors is crucial.

More than 25 years ago, the interpretation of the observations under 
the assumption of Newtonian dynamics, suggested the presence of dark 
matter in and around disk galaxies (Bosma 1978, 1981a,b; Rubin et al. 
1980, 1982, 1985; see also Rogstad \& Shostak 1972). Two relevant
(related) questions are whether the disk dominates or not the mass within 
some parts of the galaxies, and whether the dark halo is shallow or 
cuspy in the center (for recent reviews see e.g., Bosma 2002 and
Salucci \& Borriello 2001, respectively). The latter issue turned out crucial
for testing the predictions of CDM models, and it is investigated currently
by studying the inner rotation curves of observed dwarf and low surface
brightness (LSB) galaxies, which are dark matter dominated systems.

The rotation curve decomposition of {\it individual} galaxies was commonly used 
to explore the fraction of dark matter within them (e.g., Carignan
\& Freeman 1985; van Albada \& Sancisi 1986; Kent 1986). Besides of
postulating a stellar mass-to-light ratio at the observed band,
$\gamma_{\lambda}$, this method requires to assume a dark
halo mass distribution for the given galaxy.
A popular technique is to fix $\gamma_{\lambda}$ (assumed constant with radius)
at the highest possible value such that the ``luminous'' rotation curve
component does not generate a higher rotation velocity than actually
observed in the inner part, the so called ``maximum disk'' hypothesis;
the complement to the observed rotation curve is fitted to the assumed
non-hollow dark halo model (typically a pseudo-isothermal sphere). For a 
maximum disk model, the disk velocity component at its maximum dominates 
over the halo. Luminous and high surface brightness (HSB) galaxies are 
typically well described by the maximum disk model, while this model 
for low luminosity and LSB galaxies implies unrealistically high stellar
mass-to-light ratios (see references in \S 3.4).  The fractions of luminous 
and dark matter inferred with the individual velocity decomposition 
methods are model-dependent and in certain circumstances the model may converge 
to an erroneous solution 
or to a non-reasonable values for $\gamma_{\lambda}$ (Persic \& Salucci 1991).
Besides, for these methods detailed photometric and kinematic data are required.

The fractional amount of dark matter in disk galaxies have been also inferred by 
studying the kinematical properties of a set of rotation curves (Persic
\& Salucci 1998, 1990a; Salucci \& Persic 1999). The virtue of this method
is that no assumption about a particular dark matter distribution should be done.
Solely from the shape of the rotation curves and the definition of some radial 
scale connected to the luminous distribution (assumed exponential), it is possible
to infer on average the fraction of dark and luminous matter at this
scale. Persic \& Salucci  have found a dependence of 
the rotation curve slope measured at 3.2 \hd\ (\hd\ is the total disk scale radius
assumed equal to the optical scale radius) 
with the optical luminosity of the galaxy, which translates into a dependence of 
the disk-to-total mass ratio at 3.2 \hd\ on luminosity: 
M$_{\rm disk}/$M$_{\rm tot} \propto$ $L_{\rm B}^{0.4}$. The samples used by 
these authors included only HSB highly-inclined late-type spirals with 
available H$\alpha$ rotation curves and optical photometry.

A major goal of the mass modeling method based on the shape of rotation
curves was the inference of the so called ``universal rotation curve'' (Persic, 
Salucci \& Stel 1996). From about one thousand observed H$\alpha$ rotation curves
of late-type HSB spirals, synthetic rotation curves binned by luminosity intervals
were generated. Assuming an exponential disk and a halo with a velocity profile
proportional to $r/(r^2 + a^2)^{1/2}$, where $a$ is a core radius, the synthetic
rotation curves could be well described by an universal profile, resulting from 
the quadratic sum of the disk and halo velocity components. The two parameters of 
this profile are the disk-to-total velocity ratio at 3.2\hd\ and the core radius.
Persic et al. found that, in order that the synthetic rotation curves could be 
fitted with the universal profile, these parameters, in particular the former,
should be a function of luminosity. Therefore, the universal rotation curve
is a function of luminosity in the sense that as the galaxy is less luminous,
the larger is the slope of the rotation curve at 3.2\hd. A direct implication of 
this result (which was in fact the inspiration for it) is that the 
dark-to-luminous mass ratio (at 3.2 \hd) scales inversely with luminosity. 

In spite of the virtues of the rotation curve shape method 
to infer dark and luminous mass fractions, it is not free of uncertain 
assumptions (to be discussed in \S 4.2), which could bias the results. Moreover, 
other workers have shown several examples of rotation curves which deviate 
from the universal curve (Verheijen 1997; Bosma 1998), suggesting that besides 
the luminosity, other observational parameters, for example the surface 
brightness (SB), could be important, and therefore, the disk-to-total mass ratio 
could also depend on them. 
Thus, it is important to explore alternative methods based on more complete 
samples, in the sense of SBs, luminosities, and morphological types, to infer 
the dark and luminous fractions in disk galaxies.

Some of these alternative methods are based on observed global galaxy 
parameters and relationships among them rather than in the local kinematics 
(e.g., Salucci, Ashman \& Persic 1991). 
In the present paper,
we follow this kind of methods, inferring from the observations
(i) global parameters related to the amounts of luminous and dark matter
in disk galaxies, and (ii) exploring how these parameters vary with
galaxy properties. Part of the analysis presented here is closely related 
to the fundamental plane of disk galaxies. However,
a direct study of the fundamental plane will be presented elsewhere. The 
results obtained will be compared
with predictions of galaxy evolution models in the hierarchical clustering
scenario, in order to test whether disks formed inside Cold Dark Matter
(CDM) halos are realistic or not. 

 A prediction of this scenario is that the luminous-to-dark mass ratio
within the optical parts of the disks depends mainly on the disk
surface density and on the global disk(baryon)-to-halo mass fraction, 
\fd \footnote{
We define $\fd \equiv \md/\mv$, where \md\ is the total baryonic galaxy mass, 
referred some times also as the disk mass, $M_{\rm d}$, and \mv\ is the 
virial (total) halo mass.} 
(Firmani \& Avila-Reese 2000; see also Dalcanton et al. 1997; 
Mo et al. 1998). A dependence with the mass (or luminosity) is not expected, 
unless astrophysical process, like feedback, introduce a strong dependence
of \fd\ on the mass. Recent calculations
by van den Bosch (2002) show that the feedback reduces \fd\ in
low massive galaxies but there is not a significant trend with the mass. 
In fact, in massive galaxies not all the baryon fraction ends in the disk 
due to a large gas cooling time in these systems, so that both low mass
and massive systems in average incorporate 0.4-0.6 of the halo baryon
fraction to the disk (van den Bosch 2002), without any significant dependence 
on mass (i.e., \fd\ is not expected to depend on mass).

In spite of the great observational effort done in the last decades,
there is actually a few number of homogeneous enough observational works
reporting both photometric (optical and near-infrared) and kinematic 
information, including HI detection, for disk galaxies. 
In this paper, we present an extensive 
compilation from the literature for HSB and LSB disk galaxies with
photometric parameters in both the $B$ and $K$ bands, rotation curves 
or velocity line-widths and HI integral fluxes (see also Graham 2002). 
After applying an uniforming procedure, the compiled data is 
used to estimate several stellar and baryonic parameters of the disks.
The most reliable data for this aim are in the near-infrared bands ($H$ 
or $K$).
The sample and the results in this and in a forthcoming paper 
can be used as a local reference for studies of the luminous and 
dynamical properties of disks in other environments and at higher redshifts.

In \S 2, we present the compiled sample and describe the procedures
to uniformize the data and to estimate stellar and 
baryonic parameters from the observations. Sect. 3 is
devoted to the study of the ratio of disk-to-total maximum velocity at
their corresponding maxima, \vdvt. This quantity is not defined at a given
radius and gives an approximate estimate of the dark and luminous
fractions. For the kind of kinematical data we use here (equivalent
line-width instead of detailed rotation curve), \vdvt\ is the most
direct quantity to compare with models, without introducing extra
assumptions.
In \S\S 3.1 simple composite halo/disk models predictions are presented,
while in \S\S 3.2 the description and results for semi-numerical galaxy
evolution models in the \LCDM\ scenario are given. In \S\S 3.3
and 3.4, the \vdvt\ ratio of observed and model galaxies and the 
dependence of this ratio on several galaxy parameters are presented 
and compared. In the light of the results obtained, the issue of 
maximum or sub-maximum disk is discussed. In \S 4, the 
dynamical-to-baryonic mass ratio (here instead of ``total'' we use the
term ``dynamical''), \mdynmd, and the mass-to-light 
ratios (both measured at some inner radius) for models and observations,
as well as their dependences on several galaxy parameters are presented.
For the observational data, some assumptions should be done in order
to estimate these ratios; we explore how much could affect these assumptions
on our conclusions. In \S 4.2, we compare our results
with those in previous works and discuss the differences. Finally, a 
summarizing discussion is given in \S 5.

\section{The Sample}

In order to trace the baryonic and dark matter content of disk
galaxies it is important to explore a wide range of galaxy types, magnitudes,
sizes, and surface brightnesses as possible. Furthermore, high-quality 
surface photometry in the optical and near-IR pass-band as well as information 
about the rotation curve or at least the HI or H$\alpha$ line-width, and the 
total HI gas flux are needed.
After an extensive search in the literature, we surprisingly have found only a few 
(homogeneous enough) observational sources to compile a sample of
disk galaxies with the required data (see also Graham 2002). From these sources, 
three main sub-samples of HSB and LSB galaxies were collected:

(1) {\it de Jong sub-sample} (de Jong \& van der Kruit 1994, and
de Jong 1996a) reports bidimensional $B,V,R,I$ and $K$-band photometry 
for 86 undisturbed field spirals selected from the Uppsala General 
catalogue (UGC). Only galaxies classified equal or later 
than S1 or SB1 and with an inclination limit $\leq 51^{o}$
were included. This is a diameter limited sample (UGC is 
expected to be complete up to 2 arcmin red diameters) that can be 
transformed into a volume limited sample.

(2) {\it Verheijen (1997) sub-sample} (see also  Verheijen \&
Sancisi 2001) is a selection of galaxies from the Ursa Major Cluster. 
This is one of the least massive and most spiral-rich nearby cluster.
The sample is complete for galaxies later than Sab to a limiting  
apparent Zwicky magnitude mzw = 15.2, although it 
includes galaxies fainter than that cutt-off. It contains $B,V,R,J, 
K'$-band photometry and kinematic (HI observations) for HSB and LSB 
disk galaxies. The kinematic data were obtained by the authors only 
for galaxies (52) with inclinations greater than $45^{o}$ and include 
synthetic rotation curves as well as equivalent line-widths. Notice that 
although it is suspected that cluster environment affect the properties of the 
constituent galaxies, in the case of the Ursa Major, the cluster may 
be so young that it's members are more representative of a field 
population.

(3) {\it Bell et al. (2000) sub-sample,} was obtained partially by
themselves and compiled partially from the literature, mainly from
 de Blok, van der Hulst \& Bothun (1995) and de Blok, McGaugh \& van der 
Hulst (1996), with the condition to have blue central surface brightness 
$\mu_{B,o} \geq 22.5 mag arcsec^{-2}$ and diameters at the 25 mag 
$arcsec^{-2}$ isophote larger than 16$\arcsec$.
Their sample consist of 26 LSB galaxies with available bidimensional 
$B-$ and $K-$band photometry and an inclination limit $\leq 61^{o}$. The
sample is by no means complete but it is designed instead to span as 
wide a range as possible of observed LSB galaxy parameters. Notice that
although LSB galaxies follow the spatial distribution of HSB galaxies,
they tend to be more isolated from their nearest neighbors than HSB 
galaxies (c.f. Bothun et al. 1993).

The Milky Way and Andromeda galaxies were also included so that
a direct comparison of their properties w.r.t. the galaxies in our sample
could be inferred.

The galaxies from the Verheijen sub-sample and some LSB galaxies from the
Bell et al. sub-sample have measured synthetic HI rotation curves. Thus,
their maximum total velocity rotation, \vt,  is known. For the rest of 
the galaxies in our collected sample, only equivalent velocity line-widths 
are reported. In order to uniformize the data as much as possible, we used
the velocity line-width \W\ (defined at the 20\% level) for all the 
galaxies in the sample. Only a few (LSB) galaxies in the sample lack of \W\ 
but have measured rotation curves; in these cases we have
used the \vt\ inferred directly from the rotation curve. If properly
corrected, \W\ is $\approx 2\vt$ as Verheijen (1997, p. 197) have shown 
(see also Verheijen \& Sancisi 2001).
Thus, the dynamical information available for our sample will
be \vt. The radius where V$_t$ peaks, symbolized here as \rma, is not 
obviously related to a typical radius of the disk but it can be 
estimated as a multiple of \hd\ (see \S 3).
For the Verheijen sub-sample, we adopt his \W\
values, while for the de Jong sub-sample and most of the LSB galaxies
from Bell et al. sub-sample, \W\ was extracted from the Lyon Extragalactic
Database (LEDA) database\footnote{http://leda.univ-lyon1.fr}. 
The integral fluxes in the 21-cm line (used 
to calculate the neutral gas masses in galaxies) were taken from 
Verheijen \& Sancisi (2001), LEDA, and de Blok et al. (1996).

A condition imposed to our sample is that galaxies should be in a 
restricted range of inclinations ($35^{o} \leq i \leq 80^{o}$). Otherwise either 
the surface brightness (hereafter SB) profiles or the rotation velocities
are less reliable. The inclinations were calculated from the photometric 
minor-to-major semi-axis ratios (typically in a near-infrared band) reported 
in the source papers and assuming a fixed thickness parameter $q=0.2$.
We exclude from the sample galaxies with clear signs of interaction
(mainly from the Verjeihen sub-sample) and with rotation curves which are
still increasing at the last measured outer point. Only some LSB galaxies
showed this feature. Fortunately, there are published synthetic rotation 
curves for most of the LSB galaxies in our sample.
The final sample consists of 78 galaxies: 42 out of 86, 29 out of 52,
and 5 out of 23 from the de Jong, Verheijen and Bell et al. sub-samples,
respectively, plus the MW and Andromeda galaxies. The Bell et al. sample 
contains actually 26 LSB galaxies, but three of them were taken from the 
de Jong sample.  

The local distances to the galaxies were calculated using the kinematic
distance modulus given in LEDA. The value of the Hubble
constant used in this database is the same we assume here,
$H_{0}=70 \kms$Mpc$^{-1}$. Since the LSB galaxies from the Bell et al.
sub-sample are not included in LEDA, their distances were taken directly 
from Bell et al. (2000) correcting them for the H$_{0}$ value
used here. From the given distances, the corresponding redshifts were
obtained. Using these redshifts, we then calculated the luminous distances 
as:
$D_L=\frac{cz}{H_{0}}\left[ 1+\frac{1}{2}(1-q_{0})z\right]$,
where q$_0$ is the deceleration parameter and c is the speed of light.
For the cosmology used in this paper ($\Omega_m = 0.3, \Omega_{\Lambda}=
0.7$), q$_{0}=-0.55$. The above equation
is a good approximation for small redshifts.


\begin{figure}
\includegraphics[width=\columnwidth]{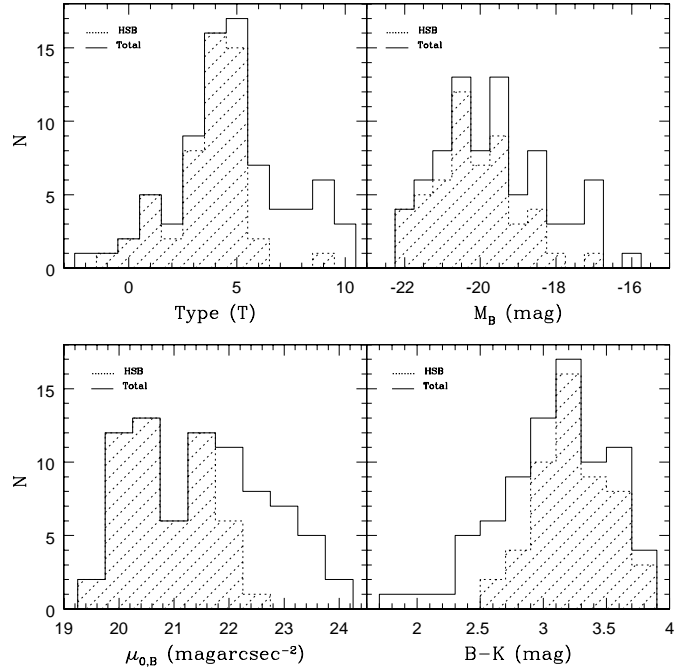}
 \caption{Number distributions of morphological types, $B-$band corrected
absolute magnitudes and SBs, and corrected $(B-K)$ colors for the
recompiled sample of 78 normal HSB and LSB disk galaxies. The contribution
of the HSB galaxies is shaded.}
\end{figure}

{\it Merits and limitations of the sample.-} The sample is not complete 
in any sense, but it is representative of the basic parameters 
required: morphological type, luminosity, SB and color. In Fig. 1, the number
distribution of morphological types, corrected $B-$band magnitudes
and SBs, and corrected $(B-K)$ colors are presented (for the corrections,
see \S 2.1). The number distribution
of types is similar to that in other samples of nearby galaxies
(c.f., Hernandez-Toledo et al. 1999; Jansen et al. 2000), with a high
peak at T = 4, 5. The distribution
of absolute magnitudes indicates that the sample is representative of 
the local galaxy luminosity function in the bright part ($M_{\rm abs}
\lesssim -19.5$) but lacks enough representation for fainter luminosities.
In fact, the faint part of the luminosity function consists mostly of dwarf 
galaxies. The physics of this group of galaxies is different from that of normal 
disk galaxies. The aim of the present paper is to investigate only the latter
ones. The SB distribution
presents a bimodality (Tully \& Verheijen 1997). The sample is not 
complete at very LSBs. This is consistent with the lack of
faint galaxies in the magnitude number distribution: very LSB galaxies are
typically dwarf galaxies.

Since we adopt the raw disk central SB and scale radius (\mul\ and \hl) 
{\it directly} from the above source samples, it is important to check that the 
the various methods used to infer these parameters do not introduce systematic 
differences in our compiled sample data.
For the de Jong sample, a two-dimensional exponential 
bulge-bar-disc decomposition was used (de Jong 1996). Bell et al (2000) used a 
one-dimensional bulge-disc decomposition (with either an exponential or an 
$r^{1/4}$ law bulge). Verheijen used a `marking the disc' fit, where the 
contribution of the bulge to the disc parameters is minimized by visually 
assessing where the contribution from the bulge component is negligible.
De Jong (1996a) showed that the one- and two-dimensional methods give comparable 
results to typically better than 20\% in terms of surface brightness 
(SB hereafter) and 10\% in terms of scalelengths, with little significant bias.

After a comparison of the de Jong and Verheijen samples in the 
L$_{\lambda}$-\hl\ diagram for the 
$B$ and $K$ bands, we found no systematic differences. Notice that
the selection criteria applied in de Jong(1996a), implies that small LSB 
galaxies are excluded. This selection effect could introduce a systematic 
difference with the Verjeihen galaxies. However, only  a few  galaxies from 
the later author fall below the selection limit of de Jong in the 
\muK-\hK\ diagram (H$_0=70\kms$Mpc$^{-1}$; this value is used throughout 
the paper). 
We also checked that the scale radii reported by these 
authors do not correlate with inclination (e.g., Giovanelli et al. 1984). 
When the bulge is not taken into account in the fitting, such a correlation 
could appear, but that is not the case for the Verheijen data.

In conclusion, we do not find significant {\it systematic} differences
in \mul\ and \hl\ among the galaxy sub-samples due to the different
profile fitting procedures used. This is consistent with de Jong (1996a)
conclusion mentioned above. In fact, the dominant source of 
error in the determination of the galaxy photometric parameters is due to 
the uncertainty in the sky level (Bell et al 2000; de Jong 1996a).

\subsection{Data Corrections}

The total magnitudes were corrected for Galactic extinction (Schlegel, 
Finkbeiner \& Davis 1998), $K-$correction (Poggianti 1997), and internal 
extinction. Since the fraction of dust seems to be larger for bigger 
galaxies according to empirical (e.g., Wang \& Heckman 1996; Tully 
et al. 1998) and theoretical (e.g., Shustov, Wiebe \& Tutukov 1997) 
arguments, we used the empirical velocity- (luminosity-) dependent 
extinction coefficients determined by Tully et al. (1998) for a large 
sample of galaxies (HSB and LSB) in clusters. The $K'$-band data from the 
Verheijen sub-sample were transformed into $K$-band adopting $K'-K=0.19
(H-K)$ (Wainscoat \& Cowie 1992) and the average $(H-K)$ color of each 
morphological type given in de Jong (1996c).

The SBs were corrected for Galactic extinction, $K-$correction,
cosmological SB dimming, and inclination (geometrical and extinction
effects). For the latter correction, we followed Verheijen (1997),
considering LSB galaxies as optically thin in all bands. The HSB 
galaxies were also considered as optically thin in the $K$-band. We have
defined LSB galaxies as those whose disk central SB in the $K-$band after
SB correction is larger than 18.5 mag/arcsec$^2$ (Verheijen 1997).

The 21 cm line-widths at the 20\% level, \W, for the Verheijen susbsample
were taken directly from  Verheijen \& Sancisi (2001). Galaxies with
\W\ taken from LEDA, were de-corrected (for instrumental corrections) to get 
the raw data (Paturel et al. 1997) and then, corrected again for broadening 
due to turbulent motions and for inclination, this time following 
Verheijen \& Sancisi (2001).

The final sample data are presented in Table 1. The name and morphological
type are given in Col. (1). The luminous distance and the galaxy inclination
are presented in Col. (2). Raw apparent magnitudes and corrected luminosities
in the $B$ and $K$ bands appear in Cols. (3) and (4), respectively. The
solar magnitudes used were $M_{B\odot}=5.48$ and $M_{K\odot}=3.41$.
The raw and corrected central SBs in the $B-$ and $K-$bands appear in Cols.
(5) and (6), while Col. (8) shows the corresponding disk scale radii
(in kpc). In Col. (7), the corrected integral and central $(B-K)$ colors
are given. Finally, Col. (9) gives the corrected line-widths at the 20\% 
level, \W, and the HI mass, 
M$_{HI}(\msun)= 2.36\times 10^{5}$D$_L^{2}$(Mpc)
$\int S_{\nu}($Jy\kms$)d\nu$.

\subsection{Composite quantities}

The stellar mass and disk central surface density, \ms\ and \sigs, are
derived from the $K-$band luminosity (which includes the bulge) and central 
SB, respectively. It seems 
that, even in the $K$ band, the stellar mass-to-light ratio, $\gamma_{K}$, 
is not constant. The application of population synthesis 
techniques to simplified galaxy evolution models for HSB galaxies show 
that $\gamma_{K}$ depends mainly on the integral color (Tinsley 1981; 
Bruzual 1983; Bell \& de Jong 2001). From Table 1 (model ``formation 
epoch with bursts (vi)'') in Bell \& de Jong (2001), we infer
\begin{equation} \label{gammaHSB}
{\rm Log}\gamma_{K}=-0.91+0.21(B-K).
\end{equation}
For low SBs this relation breaks down.
The various rotation curve decomposition models applied by Verheijen
(1997) to his sample of LSB galaxies, actually shows that $\gamma_{K}$
increases toward bluer galaxies. A linear eye fit to the $\gamma_{K}-(B-K)$
relation of his LSB galaxies, assuming a Hernquist halo model and using
his constrained decomposition method, gives 
\begin{eqnarray} \label{gammaLSB}
\gamma_{K}= 2.75-0.75(B-K'), \ \   (B-K')\lesssim 3  \\ \nonumber
\gamma_K=0.5, \ \ \ \ \ \ \ \ \ \ \ \ \ \ \ \ \ \ \ \  (B-K')>3. 
\end{eqnarray}
This result is
similar to that inferred from other decomposition methods and for the 
pseudo-isothermal halo used by Verheijen (1997; see Fig. 10 in Chapter 6). 
The scale radius of the stellar disk, \hs, is assumed here to be equal 
to the scale radius in the band $K$, \hK. Most of the stars in the disk 
are traced in this band.

The disk gas mass, \mg, is estimated as
\begin{equation} \label{mg}
\mg = 1.4 M_{\rm HI}\left[ 1+ \frac{M_{\rm H_2}}{M_{\rm HI}} \right]
\end{equation}
where the factor 1.4 takes into account helium, and M$_{\rm H_2}$ is
the molecular hydrogen mass. The M$_{H_2}/$M$_{HI}$
ratio has been found to depend on the morphological type $T$
(Young \& Knesek 1989). From the latter paper, McGaugh \& de
Blok (1997) estimated that M$_{\rm H_2}/$M$_{\rm HI} = 3.7 -0.8T + 0.043T^{2}$.
For T$ < 2$ this empirical fitting formula overestimates
strongly the gas mass in galaxies, thus we assume M$_{\rm H_2} = 0$.

The estimation of the gas and baryonic galaxy parameters is 
important to understand several structural and dynamical aspects of galaxies 
and to compare with the 
theoretical predictions. The galaxy baryonic mass is \md\ = \ms\ + \mg, and the
gas mass fraction is defined as \fg\ = \mg/\md. Unfortunately, we do not
have information about the gas surface density profiles ($\Sigma_{g,0}$ and h$_g$)
for all galaxies in our sample.  We need them in order to calculate the 
baryonic disk central surface densities and scale radii. The observed 
density profiles of neutral gas in disk galaxies are diverse
and there seems to exist a characteristic profile for each morphological type
(Broeils \& van Woerden 1994). The HI distribution in the center of 
Sb-Scd galaxies is typically flat or even has a hollow, but the H$_2$ 
distribution for these galaxies rapidly increases to the center compensating 
the HI deficiency.
For early-type galaxies, the HI profile is roughly exponential and H$_2$
is deficient. In a very rough approximation, the total gas distribution
(HI+H$_2$) in disk galaxies can be described by an exponential profile with
scale radii 2.0-3.0 times larger than the stellar scale radii in the
near-infrared bands (e.g., Olivier, Blumenthal \& Primack 1991; Marsh \& Helou
1995). We will assume that the total gas surface density follows an
exponential distribution with an scale radius 3 times \hK. Thus,
$\Sigma_{g,0}= \mg/2\pi(3\hK)^2$. The baryonic disk central surface
density is then calculated as \sigd\ = \sigs\ + $\Sigma_{g,0}$. The
corresponding baryonic disk scale radius will be
$\hd = \hK[(\sigs + 9\Sigma_{g,0})/\sigd)]^{0.5}$. In fact, our main
results in \S 3 will not be affected by including the gas disk parameters,
but some influence is expected.  This is why we tried to use
the information we have at least about the global gas content, 
by introducing  and ``effective'' gas scale radius and surface density.  

In Table 2 we present the composite quantities described above for
the sample galaxies in Table 1. Cols. (2) and (3) contain the disk stellar mass
and central surface density (\ms\ and \sigs). Cols. (4) and (5) the total disk 
gas mass and mass gas fraction  (\mg\ and \fg). Finally, Cols. (6), (7) and 
(8) list disk baryonic mass, central surface density and scale radius 
(\md, \sigd, and \hd, respectively).

\section{Disk-to-total maximum velocity ratio}

We are interested in exploring the mass amounts of luminous and dark matter
in disk galaxies and how these amounts do depend on different galaxy 
properties. Our goal is to compare the observational results with theoretical 
predictions.  As mentioned in the introduction, the analysis of 
the rotation curve shapes offers a powerful method to estimate the fraction 
of luminous and dark matter in galaxies at a given radius (3.2\hd, for example). 
The approach used here is conceptually different. In our sample, we have 
information only about the HI velocity line-width,\W, which approximately 
corresponds to twice the maximum total rotation velocity \vt\ (see introduction), 
but not about the rotation curve shapes. 
On the other hand, with the information about the disk structure that we have
(assuming exponential stellar and gas distributions), the velocity component V$_d$
due to the total disk can be calculated (eq. \ref{freeman} below), for example 
at the radius where V$_d$ attains
its maximum, 2.2\hd. Thus, without introducing further assumptions, we may define
the ratio of {\it disk} maximum velocity to {\it total} (or dynamical) maximum 
velocity, \vdvt, which can be directly compared with theoretical predictions. 

The \vdvt\ ratio is not defined at a given radius but it can be related to the 
dynamical-to-baryon (disk) mass ratio, \mdynmd, defined at the radius where the 
total rotation curve peaks, \rma. Assuming an exponential disk 
($\md\propto \sigd\hd^2$), one obtains: 
\begin{eqnarray} \label{mdynvt} 
\Bigl(\frac{\mdyn}{\md}\Bigr)_{\rm peak} \approx 
\frac{\vt^2 \rma/G}{f_L 2\pi\sigd \hd^2}
\propto \frac{\vt^2 x\hd}{f_L(x) \sigd \hd^2} \nonumber \\ 
\propto \frac{x}{f_L(x)}\Bigl(\frac{V_{\rm d,m}}{V_{\rm t,m}}\Bigr)^{-2},
\end{eqnarray}
where $\rma = x\hd$, $f_L$ is the fraction of the total disk 
mass at \rma, and \vd\ is calculated according to 
eq. (\ref{freeman}) below. So, in order to calculate \mdynmd,
one needs to define the coefficients $x$ and $f_L(x)$; they actually may change
from galaxy to galaxy, depending on the disk SB, luminosity, and halo 
density profile ($x\approx 2.2$ and $f_L\approx 0.64$ for disk dominated galaxies, 
and probably, $x>2.2$ and $f_L>0.64$ for halo dominated galaxies). 
Previous works, based on the analysis 
of rotation curves, have shown that \mdynmd\ (at 3.2\hd) has a strong dependence
on luminosity (Persic \& Salucci 1988,1990a; Persic et al. 1996). In \S 4 we 
will present results for \mdynmd. Here we concentrate on \vdvt, which is more 
directly comparable to model results than \mdynmd, and may offer a first
idea of how the fractions of dark and luminous matter in galaxies do depend
on several galaxy properties.

In order to calculate \vd\ from the observations, we
use the expression of V$_d$ for a flattened exponential disk (Freeman 1970):
\begin{equation} \label{freeman}
V_d^2(y) = 4\pi G \sigd \hd y^2 I(y),  
\end{equation}
where $y=r/(2\hd)$, $G$ is the gravitational constant, and $I(y)$ is a 
function composed of modified Bessel functions of the first and second kinds. 
The function (\ref{freeman}) has its maximum at 2.2\hd. Notice that
\sigd\ and \hd\ are not directly observed quantities (see \S 2.2). Although 
the assumptions
made to calculate them from the observational data are robust, we will 
check {\it a posteriori} whether our results are significantly dependent 
on these assumptions.

According to Sacket (1997, and more references therein), the maximum 
disk hypothesis applied to individual Sb-Sc galaxies implies a baryonic disk 
mass (including gas) which provides $(85\pm10)\%$ of the total rotation 
support at 2.2\hd\ (2.2\hK). The lower limit is occupied by galaxies with
large bulge; only in these cases the bulge contribution to \vt\ may
become significant.  From the analysis of rotation curve shapes and 
assuming the maximum disk hypothesis, Persic  \& Salucci (1990b)
reported  \md/M$_{\rm dyn}$ ratios at 3.2\hd\ increasing with L$_B$ 
from 0.3 to 0.8. This implies that (V$_d/V_t$)$_{3.2\hd}$ oscillates
between 0.57 and 0.91. The V$_d/V_t$ ratio increases from 3.2 to 2.2\hd; 
by using the rotation curve decompositions in Fig. 3 of Persic et al. (1996), 
we estimate that (V$_d/V_t$)$_{2.2\hd}$ would oscillate between 0.75
and 0.95, in excellent agreement with Sacket (1997). Although \vdvt\ is 
not formally equal to (V$_d/$V$_t$)$_{2.2\hd}$, as a reference we will say 
that the maximum hypothesis implies $\vdvt\approx 0.85\pm0.1$. 
Since the individual rotation curve decomposition methods
usually take into account the disk thickness ($q>0$), we will correct \vd\ 
as given in eq. ($\ref{freeman}$) for disk thickness. Assuming a 
minor-to-major disk axis ratio $q$ of 0.2, \vd\ becomes approximately 
5\% lower than for the infinitely thin disk case (e.g., Burlak, Gubina  
\& Tyurina 1997). 
We will apply this correction factor to the disk+halo composite mass 
models presented below as well to the observational data. 

Before studying and comparing the inferred \vdvt\ ratios with results 
from galaxy evolution models, we shall explore how this ratio is expected to
depend on different disk and halo parameters. To this purpose, simple 
disk+halo composite mass models will be used, and then they will be
compared to the complex evolutionary models.

\begin{figure}
\includegraphics[width=\columnwidth]{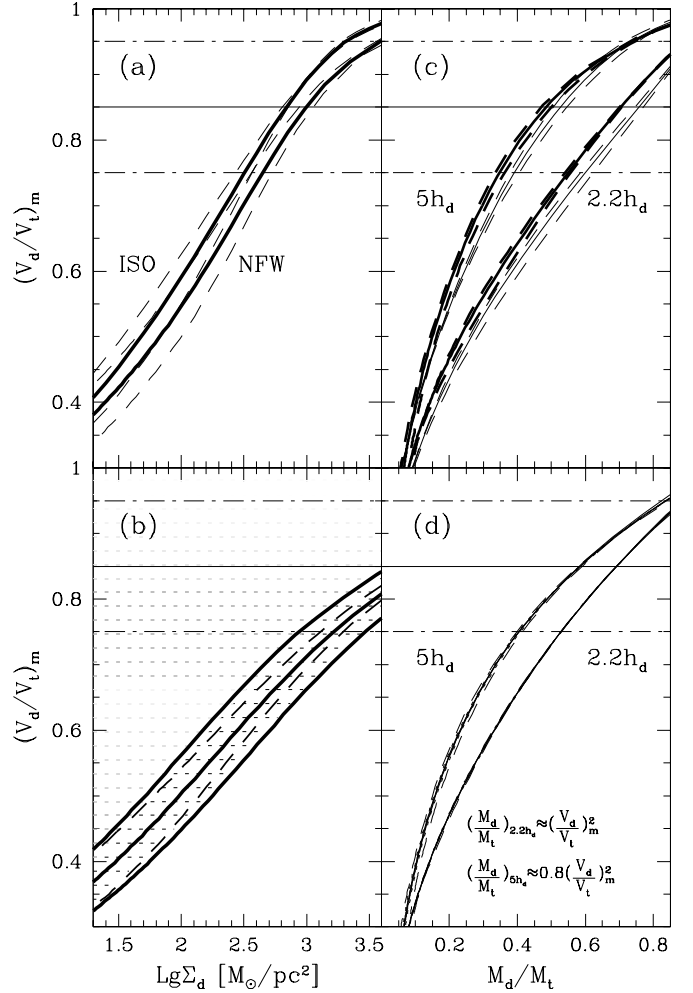}
\caption{ {\bf (a).} Correlation between \vdvt\ and \sigd\ for disks in 
centrifugal equilibrium inside a pseudo-isothermal (upper curves)
and a NFW (lower curves) halo of $\mv= 3.5\ 10^{11}\msun$. See details 
about the model parameters in the text. Solid lines are for \fd = 0.05, 
while upper and lower 
dashed lines around the solid curves are for \fd =0.08 and 0.03, 
respectively. The approximate region of maximum disk (see text) is 
bounded by the dot-dashed lines. {\bf (b).} Same as in panel (a), but 
for the NFW model, taking into account the disk gravitational dragging over
the halo. The shaded region encompasses de $\pm 1\sigma$ scatter in
concentrations around the central model. The solid upper and lower
curves are for $\mv = 3.5\ 10^{12}\msun$ and $3.5\ 10^{10}\msun$
with the corresponding average concentration parameters and \fd = 0.05. 
{\bf (c)} and {\it (d).} 
Relation between  \vdvt\ and the disk(baryon)-to-dynamical mass ratios
inside 2.2\hd\ and 5\hd\ for the same models shown in the
left panels. In panel (c) thick and thin lines are for the NFW 
and pseudo-isothermal halos, respectively. Botton in panel (d), the 
linear fittings (in Log-Log) are given.}
\end{figure}

\subsection{Composite mass models using analytical halo profiles}

Here we will analyze the rotation curve decomposition corresponding to exponential 
disks in centrifugal equilibrium inside pseudo-isothermal (e.g., de Blok \& 
McGaugh 1997) and Navarro, Frenk \& White (1997; hereafter NFW) dark halos. A 
description of how the halos and disks are constructed is given in Appendix A.  
In Fig. 2a we present \vdvt\ vs. the central disk surface density, \sigd, 
for the virial (halo) mass $\mv = 3.5\ 10^{11}\msun$
and for the two halo models. Upper and lower curves correspond to the
pseudo-isothermal and cosmological NFW halos, respectively. In the 
former case, the assumed core radius was 3.8 kpc (see Appendix A).
The solid lines are for the global baryon-to-halo mass ratio
\fd = 0.05, while dashed lines are for \fd= 0.08 (upper) and 0.03 (lower).
For the NFW halos, we have used the average concentration parameter
for a given mass taken from Eke, Navarro \& Steinmetz (2001).
The $\Lambda$CDM cosmology with $\Omega_m=0.3$, $\Omega_{\Lambda}=0.7$, 
$h=0.7$ is used here. The disk central surface density is determined
mainly by the halo spin parameter $\lambda$, but depends also on \mv\
and \fd\ (see Appendix A). Figure 2 is for a range of  $\lambda$ from 
0.015 to 0.2.   
The trend of decreasing \vdvt\ as \sigd\ decreases is
clearly seen for both halo models. Obviously,
for the pseudo-isothermal halos, the disk velocity component dominates
more than for the cuspy NFW halos.

While the pseudo-isothermal halo profile is an adhoc model introduced
to fit some observed rotation curves, the NFW profile is a theoretical
prediction for the structure of the halos in the CDM cosmogony. In the
latter case, it is important to take into account that the inner 
structure of these halos is affected by the formation of the baryonic 
galaxy, which gravitationally contracts the original halo. The usual way to
calculate this effect is by assuming a radial adiabatic invariance during the
formation of the disk (e.g., Flores et al. 1993; Mo et al. 1998; see Appendix
A). In Fig. 2b, the same curves for the NFW model presented in panel (a) are 
shown, but this time taking 
into account the halo contraction due to disk formation. Two more 
masses, $\mv= 3.5\ 10^{10}\msun$ and $ 3.5\ 10^{12}\msun$ are added 
in this panel. As one sees, when
the halo model is originally of NFW type, after disk formation, even disks with
high surface densities tend to be of sub-maximal type: \vdvt\ almost never
exceeds $\sim 0.75-0.80$\footnote{As mentioned above, the \vd\ calculated
using eq. ($\ref{freeman}$) is lowered by $5\%$ in order to take into account 
for the disk
thickness ($q=0.2$). This correction enters also in the total rotation
velocity because this last is the sum in quadrature of the disk and halo
velocity components.}. The effects of the halo contraction on the \vdvt\
ratio are non-significant only for the lowest surface density disks.
For the $\mv= 3.5\ 10^{11}M_{\odot}$ model with \fd=0.05, we also show
the scatter produced in the \vdvt-\sigd\ diagram by the different halo
concentrations. The shaded region encompasses $\pm 1\sigma$
deviations in c ($\Delta$log$c_{\rm NFW}\approx 0.18$,
Bullock et al. 2001; the lower limit corresponding to the $+1\sigma$ value).

A comparison in the \vdvt-\sigd\ plane of the pseudo-isothermal and
NFW halos (taking into account the disk gravitational drag in the last case),
shows that as \sigd\ is higher, the \vdvt\ ratio for the pseudo-isothermal models
becomes larger than that for the contracted NFW models. Thus, {\it it is at high
surface densities (SBs) where the the differences in the \vdvt\ ratio
are more sensitive to whether the halo is cuspy or shallow.} On the other
hand, neither the concentration nor the mass introduce significant
differences in \vdvt.  

For the NFW models with $\mv= 3.5\ 10^{10}M_{\odot}$ (upper thin solid line) 
and $3.5\ 10^{12}M_{\odot}$ (lower thin solid line), we also have used the 
average NFW concentrations corresponding to these masses (Eke et al. 2001)
and $\fd=0.05$. Massive systems result with lower \vdvt\ ratios than the
less massive ones, which
apparently would imply that the former are more dark matter dominated
than the latter. In fact, the average concentration for less massive
halos is higher than for the more massive ones. However, it is important to
have in mind that the parameters of the modeled disks are connected
to the scale of the halo: \hd\ and \sigd\  depend on \mv\ (see Appendix A). 
For a given $\lambda$ (constant \sigd),
$\hd\propto \vt^{\beta}$, with $\beta \approx 1.5$, i.e. shallower than 2 (Firmani
\& Avila-Reese 2000, hereafter FA00). For constant \sigd,
$\vdvt\propto \hd^{0.5}/\vt$. Therefore, $\vdvt\propto \vt^{-0.25}$,
i.e. \vdvt\ decreases with increasing \vt\ (or \mv). This effect is
partially compensated by the fact that more massive halos are less
concentrated (lower \vdvt). The final result is a small dependence
of \vdvt\ on mass. For a variation of 2 orders of magnitude in mass,
the change in \vdvt\ is around 16\% for a medium value of
$\sigd =400\msun/$pc$^{2}$ (Fig. 2b). The change of \vdvt\ for a 
$\pm 1\sigma$ in concentration is close to 12\% for the same central 
surface density.

The dependence of the \vdvt\ ratio on \fd\ is also small. Curves
with dashed lines in Fig. 2b are for the same central model
($\mv = 3.5 \ 10^{11}M_{\odot}$ and average concentration parameter) but
\fd\ from 0.08 (upper curve) to \fd = 0.03 (lower curve). The
\vdvt\ ratio does not change more than 10\% within this range
of \fd\ for $\sigd$ around $400\msun/$pc$^{2}$.

The \vdvt\ ratio is directly related to the disk and halo mass fractions.
In order to gain a more quantitative feeling of the \vdvt\ ratio, Fig. 2,
panels (c) and (d) show the corresponding disk-to-dynamical mass ratios 
at 2.2\hd\ and 5\hd\ for the same curves shown in panels (a) and (b). When 
these ratios are 0.5, the disk and the halo masses are equal at 2.2\hd\ and 
5\hd, respectively. The rotation curve peaks typically at 2.2\hd, while
5\hd\ corresponds approximately to the optical disk radius. For the 
gravitationaly contracted NFW halo model (panel d), the maximum disk case 
($\vdvt > 0.75$) implies that 53\% or more of the dynamical mass at 
2.2\hd\ should be in the disk.
At 5\hd\ this lower limit decreases down to $\approx 40\%$.
Such high values can be attained only for ultra-high surface density disks.
It is interesting to note that, for the NFW halos with contraction and for 
a large range of disk masses and surface densities, 
$\vdvt\approx (\mdynmd)_{2.2\hd}^{-1/2} = ($V$_d/$V$_t)_{2.2\hd}$.

\subsection{Self-consistent galaxy evolution models}

In previous papers (Avila-Reese et al. 1998; FA00; Avila-Reese \& 
Firmani 2000; see also van den Bosch 1998, 2000, 2002)
semi-numerical models of disk galaxy formation and evolution in 
the hierarchical $\Lambda$CDM scenario were developed. 
These models include self-consistently halo formation and evolution, 
disk star formation (SF) and feedback processes, the
gravitational dragging of the halo due to disk formation, secular
bulge formation and other evolution processes. The initial and
border conditions of the models are \mv, \fd, the halo mass aggregation
history (MAH) and the spin parameter $\lambda$. 
A short description of the main ingredients of the models is presented
in Appendix B.

Before analyzing the predictions of these models regarding the \vdvt\ ratio, it
is important to remark the following three effects. (i) The halo contraction due
to disk formation in the models was calculated for elliptical orbits, i.e.
the adiabatic invariance was not limited to only radial orbits, as was done
in the previous subsection. Particles within the virialized $\Lambda$CDM halos have
orbits with ratios of apapsis-to-periapsis around 0.15-0.40 (Ghigna et al.
1998). Therefore, the adiabatic halo contraction for the evolutionary models
presented here is of lower amplitude than in the case of assuming radial 
orbits (previous subsection). (ii) The disks formed within $\Lambda$CDM halos, 
assuming detailed angular momentum conservation, have actually
some excess in the center (cusp) w.r.t. the exponential surface density
distribution (FA00; Avila-Reese \& Firmani 2002; see also Bullock et al. 2001,
van den Bosch 2002), which give rise to a bulge formed secularly
(Avila-Reese \& Firmani 2000). The effects (i) and (ii) imply that \vd\
and the \vdvt\ ratio measured in our evolutionary models should be larger 
than in the simple
composite mass models, where the orbits were assumed radial and the disk 
surface density exponential (see previous subsection). (iii) Our model disks are
not infinitely thin; they have a vertical density and velocity structure
determined by the SF, feedback and infall mechanisms included in the modelation.
The gravitational potential of the disk is calculated taking into account
its three-dimensional structure.
A comparison of the \vd\ measured in the model galaxies with the \vd\
calculated according to eq. ($\ref{freeman}$) shows that the former is on average 
$\sim 7\%$ smaller than the latter. In fact we have corrected
(uniformly) the composite mass models (and the observational data) to take into
account the disk thickness. The $5\%$ correction in \vd\ calculated according
to Burlak et a.  (1997) is close to the 7\% measured on average in the evolutionary
models.

In Fig. 3a we present the results for 63 models grouped around three halo
masses, $\mv\approx 3.5\ 10^{10}, 3.5\ 10^{11},$ and $3.5\ 10^{12} M_{\odot}$
(triangles, squares and circles, respectively) with MAHs and $\lambda$'s
obtained randomly from their corresponding (cosmological) statistical 
distributions, and $\fd = 0.05$ (FA00; Appendix B).
The solid curves are the same curves plotted in Fig. 3b for the three masses,
the average concentration parameters, and \fd = 0.05. The evolutionary
models seem to lie slightly higher than the corresponding
composite mass models in the \vdvt\ vs. \sigd\ plane, in particular at ultra-high
surface densities, where the disk cusps are more pronounced. Although an attempt
to introduce a correction due to disk thickness has been done for the
composite mass models, it is important to keep in mind the other two effects
mentioned above as well several evolution processes included in
the modelation. Nevertheless, the simple disk+halo composite mass models offer 
a reasonable description of results from the evolutionary models and deserve 
to be included. 
These models can be easily reproduced by other workers in the field 
interested in using these kind of theoretical results.

\begin{figure}
\includegraphics[width=\columnwidth]{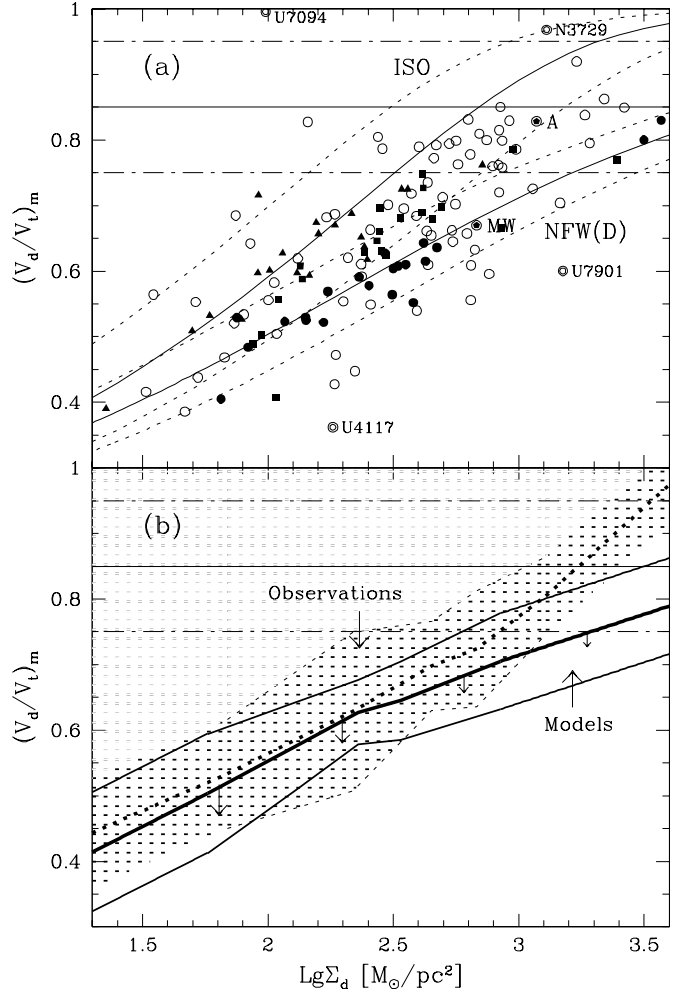}
\caption[]{{\bf (a).} Same as in Fig. 2 but for observations
(empty circles) and evolutionary model results (solid triangles,
squares, and circles corresponding to $\mv = 3.5\ 10^{10}\msun,
3.5\ 10^{11}\msun$, and $3.5\ 10^{12}\msun$, respectively). The models
are for \fd = 0.05. Four galaxies, which are outliers in this plot, as
well as the Milky Way and Andromeda galaxies are indicated by their names.
For indication, the composite mass models presented in Fig. 2 for the 
pseudo-isothermal halo (upper curves) and the NFW halo with gravitational 
drag (lower curves)
are also shown. The solid lines are for $\mv = 3.5\ 10^{11}\msun$ while
the dashed ones are for $3.5\ 10^{10}\msun$ (upper curves) and
$3.5\ 10^{12}\msun$ (lower curves).
{\bf (b).} Means per \sigd\ bins and their corresponding standard deviations
for the observational data (thick dashed line and shaded area) and the model
results (thick solid line and empty frame) showed in panel (a). The
arrows from the thick curve are an estimate of how much the models would 
decrease if \fd\ is decreased to 0.03.}
\end{figure}

The model \vdvt\ ratio is determined mainly by the
spin parameter $\lambda$ (FA00), therefore \vdvt\
correlates mainly with \sigd\ as seen in Fig. 3a.
The models show also a slight dependence of the \vdvt\
ratio on mass (see the segregation of triangles, squares and
circles in Fig. 3a). The reason of this dependence has been
already explained in the previous subsection. For a given mass,
the scatter seen in Fig. 3a is due to the scatter in the halo
structures. This scatter has its origin in the dispersion of the
MAHs (Avila-Reese et al. 1998; Wechsler et al. 2002). For perfect
NFW halos, this scatter is completely described by the scatter in
the NFW concentrations. Thus,  {\it more concentrated halos 
and more massive galaxies tend to have lower \vdvt\ ratios for a given
\sigd.} The differences are actually small.

\subsection{Observational data}

In Fig. 3a we plot the \vdvt\ ratio vs. \sigd\ inferred
from the compiled observations in \S 2. The scatter is large and
before comparing to models, we shall explore whether
the observed correlation is significant and direct. In Fig. 3a, four galaxies
have been identified as outliers (shown with their names in the
plots). These galaxies were excluded from any statistical analysis.

In a first exploratory phase, we look for any correlation of the \vdvt\ 
ratio (Log\vdvt) with several galaxy properties (independent
variables): Log\sigd, Log\hd\ (or Log\md), $(B-K)$ color, morphological type T, 
and Log\fg.
We applied backward step-wise multiple linear regressions, where F-tests are
used to compute the significance of each independent. The elimination of
non-significant variables leads us to a bidimensional linear correlation,
where the two more significant independents are \sigd\ and T, although it could
also be \sigd\ and $(B-K)$. The correlation is:
\begin{equation} \label{vdvt1}
\vdvt \propto \sigd^{0.20\pm 0.02}T^{0.01\pm 0.003},
\end{equation}
with the intercept equal to $10^{-0.72\pm 0.06}$ and an adjusted
correlation coefficient R=0.79 (\sigd\ is in unities of \msun pc$^{-2}$). 
The dominant correlation is with \sigd. 
Notice that although a direct linear regression renders
$\vdvt \propto \sigd^{0.14\pm0.02}$, with an intercept 
of $10^{-0.54\pm 0.04}$ and R=0.75, as is seen in eq. ($\ref{vdvt1}$),
the correlation significance is improved slightly introducing T, although
the slope for T is negligible. The reason for this is that \sigd\ and T are not
completely independent among them; \sigd\ (anti)correlates slightly with T.

The most important result of our analysis is that {\it \vdvt\ depends mainly on
\sigd.} In the past, the conclusions of various observational studies
were that the disk and halo mass fractions in the optical parts of disk
galaxies depend on the SB and/or luminosity (or scale radius). Applying
a direct linear multidimensional regression to \vdvt\ with
\sigd\ and \hd\ as independents, we find that:
\begin{equation}  \label{vdvt2}
\vdvt \propto \sigd^{0.12\pm0.02} \hd^{-0.07\pm0.03},
\end{equation}
with R=0.75. The large 
uncertainty in the slope of \hd\ shows that the significance of this 
variable is small, in agreement with the result of the F-test.

We arrive to similar conclusions by making use of the fundamental plane (FP)
of disk galaxies. An extensive analysis of the luminous and baryonic FPs for
our sample will be presented elsewhere. For the baryonic case and applying 
the direct linear regression fitting, we obtain that
$\vt\propto \sigd^{0.37\pm 0.02} \hd^{0.57\pm 0.03}$
with R=0.95. Combining this relation with the definition of \vdvt\ 
($\propto [\sigd \hd]^{1/2}/\vt$), eq. ($\ref{vdvt2}$) is recuperated.

We explore now any dependence of the \vdvt\ ratio on more direct
observational quantities. We repeat the step-wise multiple regressions using
now as independents Log\sigK\ and Log\hK\ instead of Log\sigd\ and
Log\hd, besides $(B-K)$, T, and Log\fg. The statistical analysis shows
again that only two independents are significant: \sigK\ and $(B-K)$ or
\sigK\ and T. The most significant correlation is:
\begin{equation}  \label{vdvt3} 
\vdvt\propto \sigK^{0.16\pm0.02} (B-K)^{-0.08\pm0.02},
\end{equation} 
with R=0.76. The regressions for \sigK\ and T, and \sigK\ and \hK\ give:
$\vdvt\propto \sigK^{0.17\pm 0.02} T^{0.014\pm 0.004}$ (R=0.75), and 
$\vdvt\propto \sigK^{0.11\pm0.01}\hK^{-0.09\pm0.03}$ (R=0.72), respectively.
In the latter case, note the large uncertainty in the slope of \hd.
The trends seen above (eq. $\ref{vdvt2}$) with the composite quantities 
\sigd\ and \hd\ remain approximately similar when using the observables 
\sigK\ and \hK.  We have also repeated the statistical analysis, this
time neglecting the gas contribution, that is, \vd\ was calculated only from 
\sigs\ and \hs (= \hK). In this case, the correlation of \vdvt\ with the surface 
density becomes even slightly steeper than in the case including the disk gas 
parameters.
Therefore, our conclusion about the dependence of \vdvt\ 
on the disk surface density and its small dependence on the scale radius
is robust w.r.t. the the assumptions made to calculate \sigd\ and \hd\ 
from the observational data (see \S 2.2): the stellar M/L ratio $\gamma_K$,
and the disk gas parameters, inferred from the total gas mass.

It is important to remark that if the galaxy sample is  
restricted to a relatively small range of luminosities and/or SBs, then 
apparent anti-correlations between \vdvt\ and \hd, and luminosity or mass appear
(in Figs. 7 and 8 we show related effects for the mass-to-luminosity
ratio). 
Therefore, for samples limited in luminosity and/or SB, the \vdvt\ ratio 
would seem to be larger for smaller galaxies. In fact, even for a
complete sample, a slight trend in this direction appears, 
but it is probably due to the dependence of \vdvt\ on \sigd\ and the fact
that smaller galaxies are typically also of lower SB.

\subsection{Observations and models: maximum disk?}

Observations and models show that the \vdvt\ ratio changes among
galaxies mainly depending on their surface densities (brightnesses):
it decreases monotonically as the galaxy is of lower surface
density (Fig. 3a). The horizontal lines in Fig. 3a show the range
of maximum disk according to Sacket (1997), (V$_d/$V$_t$)$_{2.2\hd}= 
0.85\pm 0.10$, which for the \vdvt\ ratio should be considered only as 
an approximation for low SBs).
Notice that some fraction of the {\it observed galaxies of ultra-high and
high SB are described by maximum or marginally maximum disks, while LSB
galaxies tend to be sub-maximum disks}. Indeed
several detailed observational studies of rotation curve shapes and
SB profiles for HSB galaxies show that they tend to
be maximum disks (e.g., Corsini et al. 1998; Salucci \& Persic 1999;
Palunas \& Williams 2000).
Theoretical arguments, including the swing amplifier constraints or
bar slowing down due to dynamical friction also suggest a dominion of
luminous matter in the center of HSB
galaxies (see Bosma 2002 for a review and references therein). On the other
hand, for LSB galaxies, a maximum disk solution demands too high stellar M/L
ratios, which for reasonable stellar population models imply non-realistic
color indexes for the disks (de Blok et al. 1997,2001). The results
in Fig. 3 complement all these previous works, {\it unifying HSB and LSB
galaxies in a continuity of decreasing  luminous-to-total mass
ratio within the optical part of the galaxy as the disk SB decreases.}
Nevertheless, this ratio varies significantly even for a fixed SB.
Below we explore whether other galaxy properties play any significant role 
in defining the luminous-to-dark matter ratio.

From the theoretical side, for disks formed within
$\Lambda$CDM halos assuming \fd = 0.05, only those with the
highest surface densities could be described by the maximum disk solution.
Galaxy models with moderate \sigd\ ($\sim 400\msun $pc$^{-2}$) have 
sub-maximal disks, with $\vdvt \approx 0.66\pm 0.06$ (for the NFW
halos, $\vdvt \approx$ V$_d/$V$_t$)$_{2.2\hd}$ almost ever, as shown
in Fig. 2d). If \fd\ is decreased 
to the more realistic values of 0.03-0.02 (Avila-Reese, Firmani
\& Zavala 2002), then models result with even lower \vdvt\ ratios. The arrows 
in panel (b) of  Fig. 3 indicate how much the
\vdvt\ ratio of the models would decrease if instead of \fd = 0.05,
\fd = 0.03 is used. In this panel the means per \sigd\ bins and
their corresponding standard deviations are also plotted. The shaded region 
is for observations, while the region within solid lines is for the models.
The thick dotted and solid lines connect the bin means of both groups, 
respectively.

Observations and models show a large and roughly similar scatter in
the \vdvt-\sigd\ plot (Fig. 3).
For the former, part of this scatter is due to the combination of observational
uncertainties in the measurements of \vt, \sigK, \hK, and M$_{\rm HI}$ propagated
in the calculations of \vdvt\ and  \sigd\ (see Graham 2002 for
uncertainty estimates for some of the observational data). The residuals of the
observational \vdvt-\sigd\ relation do not correlate in a significant way with 
any other galaxy parameter, like \fg, \md, \vt, morphological type, integral
$(B-K)$ color. Nevertheless, we find {\it some trend for redder, earlier 
type, and more massive galaxies to have on average smaller \vdvt\
 ratios}. 
The trends with the type and color would increase slightly if one takes
into account that, as the galaxy is of earlier type and redder, the thicker
its disk is. The thicker the disk, the lower the \vdvt\ ratio is. Notice that 
the observations have been uniformly corrected for disk thickness
for all the galaxies (5\% in \vdvt). The trends just mentioned can be
explained by our models. 

For the models, the scatter in the \vdvt-\sigd\ relation for \fd\ constant
is produced mainly by variations in the halo structure (concentration) and
the mass (see Fig. 2b).  Since more concentrated halos are the product of
earlier mass assembling (Avila-Reese et al. 1998), 
the disks formed within these halos with a gas accretion rate proportional
to the MAH, will be redder. Thus, for a given \sigd, redder disks are
expected to have lower \vdvt\ ratios. The weak correlation with the mass
was already explained in \S\S 3.1. These effects can be interpreted as
weak deviations from homology and show a connection between halo structure 
and luminous evolution. Unfortunately, as mentioned above, the
observational uncertainty is too large to allow a direct comparison
of models and observations regarding the correlations of the residuals.
Nevertheless, at least the trends seem to be in the same direction, showing
the good predictive capability of the galaxy evolutionary models of FA2000.

The comparison of observations and models in panels (a) and (b) of Fig. 3
shows some agreement among them, but on average models fall below
observations. For $\fd = 0.03$, the difference increases (see the arrows 
in Fig. 3b). In panel (a) we also reproduce the line corresponding to the 
pseudo-isothermal
model with \fd = 0.05 ($\mv=3.5\ 10^{11}\msun$, r$_c$=3.9 kpc)
presented in Fig. 3a, and add results for two other masses:
$\mv=3.5\ 10^{10}\msun$ (r$_c$=1.9 kpc, upper line) and
$\mv=3.5\ 10^{12}\msun$ (r$_c$=7.7 kpc, lower line); the
core radii are in accordance to inferences by Firmani et al. (2001), see \S 3.1.
As was noted previously (Figs. 2a and 2b), the major difference between cuspy,
gravitationally contracted, NFW and pseudo-isothermal halos in the
\vdvt-\sigd\ plane is at high surface densities. The observations lie
actually in between both halo models, specially at high surface densities.

The galaxies formed within $\Lambda$CDM halos seem to be somewhat less 
dominated by disk than observations suggest, specially when
\fd=0.03. The average difference in this case would be roughly of 
$10\%$.  On the other hand, a pseudo-isothermal halo yields a \vdvt\ ratio
larger than observational inferences for HSB galaxies (and a steeper
increase of the \vdvt\ ratio with \sigd). These differences might
decrease by using smaller core radii than assumed here or by introducing the
disk gravitational drag over the pseudo-isothermal halo. The fact that
the \vdvt\ ratio continues increasing with \sigd\ for ultra-HSB observed 
galaxies could be pointing to that either the disk gravitational drag on the
halo is not as efficient as in the models (the gravitational drag is the 
responsible that \vdvt\ tends to be constant as \sigd\ increases to high 
values) or to 
that some physical process is expanding continuously the halo core 
(self-interacting CDM for instance).

We conclude that the predictions of the $\Lambda$CDM scenario regarding the
amounts of dark and luminous matter in disk galaxies are in marginal
agreement with observations, specially if $\fd < 0.05$.
However, the way in which the amounts of dark and
luminous matter change as different galaxy properties vary (\sigd\ in
particular) is in good agreement with observations. It seems that minor
modifications to the inner structure of the $\Lambda$CDM halos is enough
to conciliate models with observations.
In FA00 we have shown that the introduction of soft cores in agreement
with inferences from the rotation curves of LSB galaxies, produce a
$\sim 10\%$ increase of the \vdvt\ ratio (on average) for HSB galaxies.
These models would be in better agreement with observations in Fig. 3.

\section{Dynamical-to-baryonic mass and mass-to-light ratios}

 In the previous Sect. we have investigated the \vdvt\ ratio from
observed and model galaxies and determined how it depends on
different galaxy properties. As shown in Fig. 2, panels c) and d),
\vdvt\ seems to be good tracer of the amounts of luminous and dark 
matter in disk galaxies. In this section we will study directly
the dynamical-to-baryonic and dynamical-to-stellar mass
ratios, \mdynmd\ and \mdynms, as well as the mass-to-light ratios.
The latter is the ratio most directly obtained from observations and
it has been determined previously in the literature, mainly in the
$B-$band (e.g., Salucci, Ashman \& Persic 1991; Karachentsev 1991;
Roberts \& Haynes 1994; Graham 2002). We remark that L$_K$ and L$_B$ 
refer to the {\it total} galaxy luminosities, i.e. disk+bulge.
Therefore, \ms\ and \md\ are the total disk+bulge stellar and baryonic
masses, respectively.  In general, while the \vdvt\ ratio refers to the disk 
contribution (\vd\ has been calculated from the {\it disk} SB 
profile (eq. [$\ref{freeman}$]), the \mdynmd\ ratio refers to the total 
galaxy contribution (disk + bulge). Thus, from the 
observational point of view, the information contained in 
\mdynmd\ is slightly different to that contained in \vdvt.

In \S 3, eq. (\ref{mdynvt}), we have shown the theoretical relation between 
our observational quantity \vdvt\
and \mdynmd\ defined at \rma, the radius where the rotation
curve peaks. For disk dominated galaxies, the radii where the disk 
and total velocities peak are approximately the same, at 2.2\hd\
($x=2.2$), and the disk mass fraction at this radius is $f_L=0.64$.
For halo dominated galaxies, the departure of $x$ and $f_L$ from 
2.2 and 0.64 will depend mainly on the halo mass distribution;
in any case, both will increase, in such a way that the ratio 
$x/f_L$ will only slightly change (increase). For example,
for NFW halos, $x/f_L$ remains almost constant ($\approx 2.2/0.64$)
for almost any case (see Fig. 2d). Thus, we may estimate approximately
the \mdynmd\ ratio at 2.2\hd\ by multyplying $\mdynmd\equiv (\vt^2\hd/G)/\md$ 
by 2.2/0.64=3.44. We may also estimate \mdynmd\ at
$5\hd\ = 5 \rma/x$ (roughly the optical radius of disks). For this,
we have to assume that the rotation curve remains flat from its
maximum up to this radius, and \mdyn\ should be multiplied by 5;
for an exponential disk, the mass at 5\hd\ is close to the total
disk mass (99\%). In fact, the outer slope of the rotation curves
changes, mainly depending on the disk surface density and luminosity. While some
LSB and low-luminosity galaxies may have still increasing rotation 
curves at $\sim 5\hd$, HSB and luminous galaxies may present decreasing 
shapes at this radius (e.g., Casertano \& van Gorkom 1991; Persic
et al. 1996; Verheijen 1997). Galaxies with 
increasing rotation curves at the last measured point
were excluded from our sample. The highest
SB and luminosity galaxies may have on average a rotation velocity at 5\hd\
smaller than \vt\ by a $10-20\%$. Therefore, the \mdynmd\ ratio at 5\hd\ presented
below could be overestimated by this factor for the highest SB and luminosity
galaxies.

\subsection{Results}

In panels (a), (b), and (c) of Fig. 4 we plot $(\vt^2\hd/G)/\md$ vs. \sigd,
$(\vt^2\hd/G)/\ms$ vs. \sigs, and $(\vt^2\hd/G)$/L$_K$ vs \sigK, respectively.
The axis to the right of each panel was shifted up by a factor 
of 5 to resemble approximately \mdynmd, \mdynms, and \mdynlK\ at 5 \hd, 
respectively. To obtain these ratios at 2.2 \hd, the shift should be approximately
by a factor of 3.44. Similar to \vdvt, the
\mdynmd\ ratio presents a dependence on \sigd, with a poor correlation with
\hd. Applying a direct multidimensional linear regression
to \mdynmd\ as a function of \sigd\ and \hd\ (Log-Log), we arrive to the 
same conclusion of the previous section: {\it the dynamical-to-baryonic mass
ratio in disk galaxies depends mainly on \sigd\ but weakly
on scale (\hd\ or \md)}. The correlation obtained from the observations is
\begin{equation} \label{mdyn1}
\mdynmd \propto \sigd^{-0.31\pm 0.03}\hd^{-0.05\pm 0.05},
\end{equation}
with R=0.8. This is consistent with eq. ($\ref{vdvt2}$) 
if $\mdynmd\propto \vdvt^{-2}$.

The direct linear regression of Log(\mdynmd) with Log\sigd\ gives
$\mdynmd = \alpha 10^{0.63\pm 0.10} \sigd^{-0.31\pm 0.04}$ (R=0.74), 
with $\alpha = 5$ 
(solid line in Fig. 4) and  $\alpha = 2.2$ for this ratio defined 
at 5 and 2.2 \hd, respectively. The ratio of baryon to dark matter,
$\md/\mh$, is related to \mdynmd\ by:  $\md/\mh = 1/(\mdynmd -1)$.
Thus, on average the $(\md/\mh)_{\rm 2.2\hd}$ ratio for ultra-HSB 
($\sigd\approx 2000\msun$pc$^{-2}$), moderate-HSB 
($\sigd\approx 400\msun$pc$^{-2}$), and very-LSB 
($\sigd\approx 50\msun$pc$^{-2}$) galaxies are 2.5, 0.8 and 0.25,
respectively. For the MW and M31 galaxies, $(\md/\mh)_{\rm 2.2} = 0.74$ 
and 1.34, respectively.

The anti-correlation of \mdynmd\ with the surface density becomes steeper 
(and tighter)
when one passes from the baryonic to the stellar disk or the $K-$band
(panels b and c). A direct multidimensional lineal regression shows that
$\mdynms \propto \sigs^{-0.51\pm 0.04}\hK^{-0.13\pm 0.06}$ and
$\mdynlK \propto \sigK^{-0.56\pm 0.03}\hK^{-0.16\pm 0.06}$ (R=0.86 and
0.89, respectively). This change is because
galaxies with lower surface densities have typically higher gas fractions
than the higher surface density ones. A contrary effect happens when
one passes from stellar masses and densities to the $B-$band luminosities
and SBs: lower surface density galaxies are typically
more luminous in the $B-$band (bluer) than the higher surface density 
galaxies with the same masses.
Therefore, the  \mdynl\ vs. \sigB\ relation (Fig. 5a) becomes shallower
and noisier than the \mdynms\ vs. \sigs\ relation. Nevertheless, we find
that {\it \mdynl\ still (anti)correlates significantly with \sigB,
and does not correlate with \hB} (Fig. 6a). A multidimensional linear
regression shows that 
\begin{equation} \label{mdynlb}
\mdynl\propto \sigB^{-0.37\pm0.04}\hB^{0.03\pm0.06}
\end{equation}
with R=0.77.

\begin{figure}
\includegraphics[width=\columnwidth]{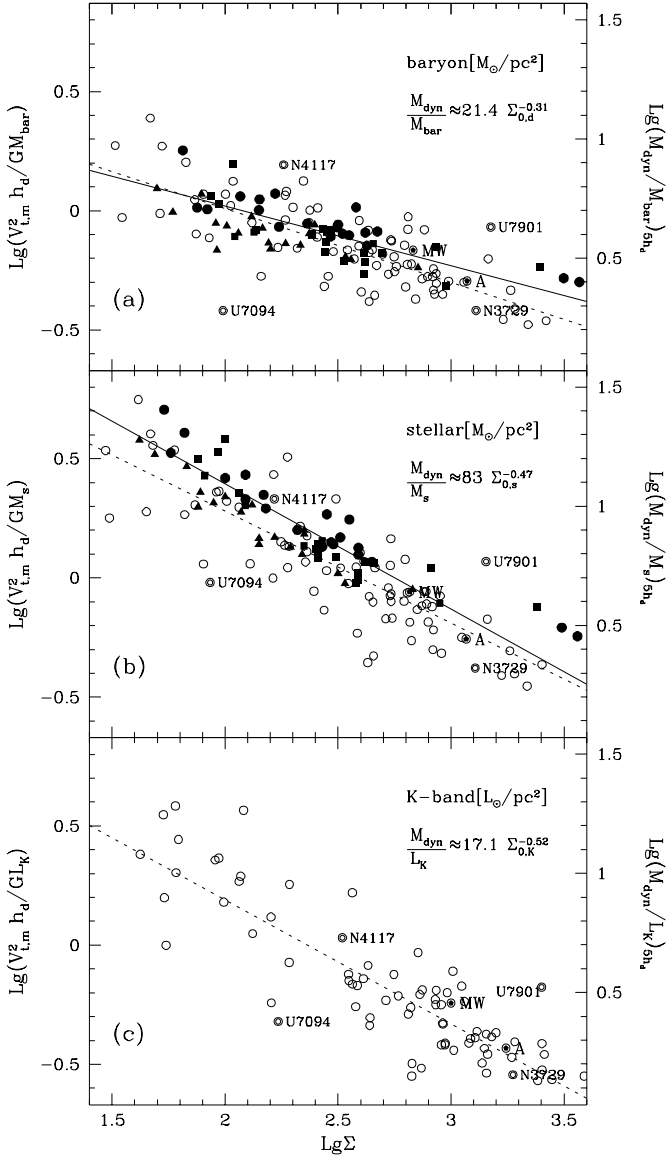}
\caption{{\bf (a).} Dynamical-to-baryonic mass
ratio vs. the baryonic surface density for observations (empty circles) 
and models (same symbols as in Fig. 3). The left axis is the quantity 
inferred directly from the observational data, while the right axis is 
this quantity multiplied by 5; in the assumption that the rotation curve 
remains constant up to 5\hd, this corresponds to \mdynmd\ at 5\hd. Solid 
and dotted lines are the linear regressions for models and observations, 
respectively. For observations, the fitting (at 5\hd) is showed in the 
upper right. {\bf (b).} Same as in panel (a) but for the stellar mass and
surface density. {\bf (c).} Same as in panel (a) but for the $K-$band
luminosity and SB. Only observations are showed in this panel.}
\end{figure}

It should be noticed that if the sample is strongly restricted in 
magnitudes, a spurious correlation of \mdynl\
with \hB\ appears. For instance, if the data in Fig. 6a (\mdynl\ vs.\hB) 
are sorted into three groups according to their $B-$band magnitudes, then
each group show an apparent correlation. We also have calculated
the residuals of the \mdynl-\sigB\  and  \hB-\sigB\
relations. There is not any dependence among them. Contrary to this, 
and using the FP, Graham (2002) arrived to the 
conclusion that \mdynl\ does not depend on \sigB\ but with 
\hB. In \S4.2 we discuss the reasons of this apparent difference.

The correlations we have found for \mdynlK\ and \sigK, and for
\mdynl\ and \sigB\ imply the existence of a dependence of L$_B$/L$_K$
on both \sigB/\sigK\ and \sigK. With some scatter, this dependence can
be written as
\begin{equation} \label{B-K1}
(B-K) \approx 0.36(B-K)_0 - 0.19\muK,
\end{equation}
which agrees with a direct
comparison of these observable parameters. The disk center is typically
redder than the average disk color and the difference increases for
redder galaxies (e.g., de Jong 1996c; Verheijen 1997).

In Fig. 7a we explore whether the \mdynmd\ ratio (defined at 5\hd)
correlates with the integral $(B-K)$ color. Apparently there is no
significant direct correlation, but a weak anti-correlation (see also 
Graham 2002). This is mostly because redder galaxies are commonly of higher 
SB. As was seen above, galaxies with higher SB are less dark matter dominated
than galaxies with lower SB. However, for a given SB, one expects
that redder galaxies (assembled earlier) are more dark matter dominated, at
a given SB, than the bluer galaxies  (Avila-Reese \& Firmani 2000). This is
because dark halos assembled by an early collapse are more  
concentrated than halos assembled by extended accretion.
In Fig. 7b, results from the same evolutionary models presented in 
\S 3 are shown in terms of the $(B-V)$ color. 
The data were sorted into three groups according to their disk surface 
densities. For the lower surface density models, \mdynmd\ correlates 
with the color as expected. For higher  surface densities the correlation 
tends to disappear because of the effect of the surface density and the
``saturation'' of the dark-to-baryonic mass ratio at high SBs. This is due
to the disk gravitational drag over the halos (otherwise, the bluer HSB
models could have lower \mdynmd\ ratios). In panel (a) the observational
data has also been sorted into the same three groups according to their
disk surface densities as in panel (b). The trend for the low surface
density galaxies is similar than that for the models.  At high surface 
densities observations also agree roughly with models. However,
several of the highest surface density and reddest galaxies appear 
with lower \mdynmd\ ratios than models.
These galaxies, having the most dominant disks and (probably)
the more concentrated halos, reveal that dark matter is not as dominant
as in the case of disks formed within NFW halos.

\begin{figure}
\includegraphics[width=\columnwidth]{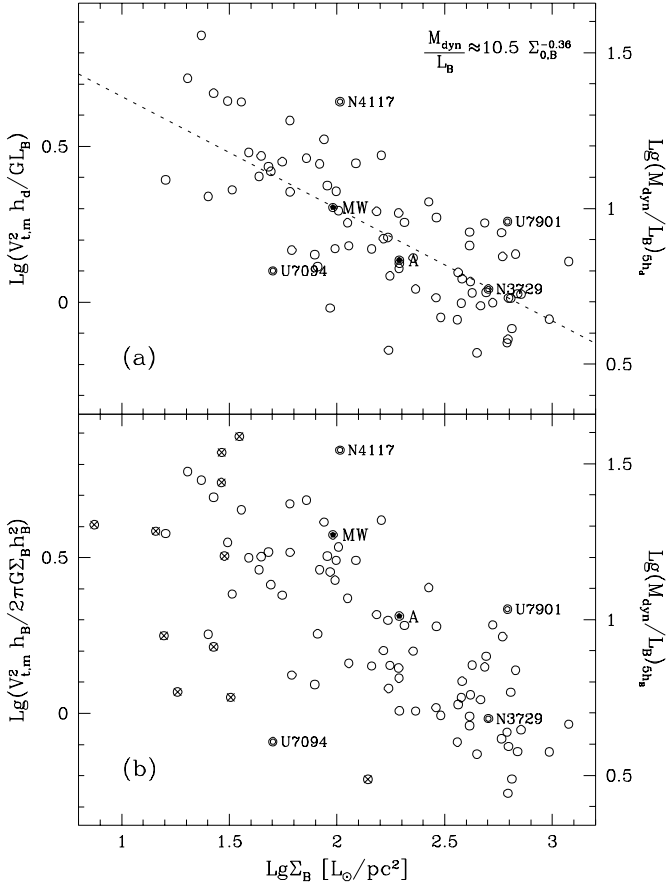}
\caption{{\bf (a).} Same as Fig. 4, but for  $B-$band luminosity and
SB, and only for observational data. {\bf (b).} Same as panel (a)
but using the definitions of Graham (2002; see text) and including 
some LSB galaxies with large bulges also considered by Graham (circles
with crosses). Notice how the scatter in the correlation seen in panel 
(a) increases in panel (b).}
\end{figure}

\begin{figure}
\includegraphics[width=\columnwidth]{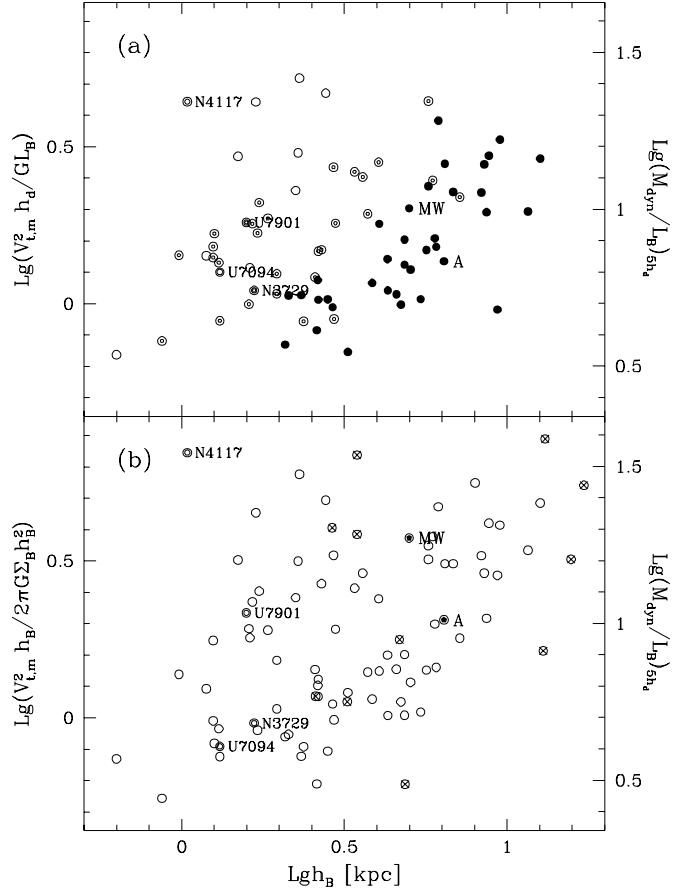}
\caption{{\bf (a).} Same as in Fig. 5a but
for \hB\ instead of \sigB\ in the x-axis. The data are sorted into
three groups according to their B-band magnitudes. The
corresponding ranges are: $4.0\ 10^{8}-2.5\
10^{9}$L$_{B\odot}$ (open symbols), $2.5\ 10^{9}-1.6\
10^{10}$L$_{B\odot}$ (targeted symbols), and $1.6\
10^{10}-1.0\ 10^{11}$L$_{B\odot}$ (solid symbols).
Although for all the data is not significant, for the data sorted
in luminosity ranges, a correlation between the mass-to-light ratio 
and \hB\ appears. {\bf (b).} Same as in Fig. 5b
but for \hB\ instead of \sigB\ in the x-axis. A correlation appears
for all the data in this panel (calculated according to Graham 2002).
}
\end{figure}

In Fig. 4, panels (a) and (b), the data from the evolutionary models  
are also included (solid
symbols; but notice that there are no model predictions in the $K$ band).
Solid and dashed lines are direct linear regressions to the observational
and model data, respectively. As in the case of
the \vdvt\ ratio, models agree marginally with observations, yielding
on average higher \mdynmd\ and \mdynms\ ratios than
observations, specially for the highest surface density galaxies.
For example, for the groups of ultra-HSB and moderate-HSB galaxies 
mentioned above, the model (\md/\mh)$_{\rm 2.2\hd}$ ratios are on average
1.4 and 0.6 (with a large scatter), which means 
approximately 1.8 and 1.25 times more dark matter dominated than
observations, respectively. For the LSB galaxies, the 
ratios are roughly similar. The differences between observations and models 
increase if \fd\ is fixed to values smaller than 0.05. The halos of
observed galaxies have on average less dark matter content in
the center than the \LCDM\ model suggests. It should be
also noted that the differences between models and observations
decrease when passing from stellar to baryonic (stars+gas) parameters.
This is because models have slightly higher gas fractions
than observations (FA00). In our models all the gas is cold, while 
the observations suggest that some fraction of the gas may 
be ionized in the outer disks. Therefore, this gas in not detected
in the HI observations.

\subsection{Comparison with previous works}

Using the tilt of the optimal fundamental plane (FP) of disk galaxies,
Graham (2002) concluded that \mdynl\ depends on \hB\ rather than
on \sigB. Since we share almost the same observational sources,
our results broadly agree with those of Graham (2002) if we follow
his FP analysis. In this indirect way to estimate the dependence of
\mdynl\ on \sigB\ and \hB, the multidimensional regression is applied
not to \mdynl, but to \vt\ (the FP). This is one of the reasons behind
the different conclusions achieved in Graham's paper.
From his data, one sees that if the
mass-to-light ratio dependence on \sigB\ and \hB\ is determined
directly (``kinematic'' M/L ratio) instead of through
the FP (the ``photometric'' M/L ratio),
then the dependences on \sigB\ and \hB\ increases and decreases
slightly, respectively (see his Fig. 4), and the result approaches to
what we have reported in the previous subsection.

Other reasons to explain the difference in results
are the way to calculate \mdyn\ and L$_B$.  Graham (2002) uses the
disk scale radius in the current band while we define a unique physical
\mdyn, using the baryonic disk scale radius \hd. Rather than the 
measured L$_B$, Graham uses $\sigB\hB^2$. Figures 5 and 6, panel (b), show
our data using the same definitions as in Graham (2002). We also
include in these plots LSB galaxies with large bulges from the
sample of Beijersbergen (1999) and de Blok et al. (1998) (circles
with crosses; for these galaxies there is available information
only in the $B-$band), which were used by Graham. It is
clearly seen that the dependence of \mdynl\ on \sigB\ becomes
weaker in panel (b) of Fig. 5 than in panel (a). On the other
hand, as in Graham (2002), some correlation with \hB\ appears now
(compare panels a and b of Fig. 6). Thus, our results agree
broadly with those of Graham (2002) when we apply to the data his
technique and definitions, and when we include the peculiar
sub-sample of LSB galaxies with large bulges. 

Nevertheless, we
think that our analysis and the conclusion in \S 4.1 
that \mdynl\ anti-correlates with \sigB\ and does not depend on \hB\ or 
L$_B$ is physically correct. If we use the result of Graham (2002), it is
easy to show that $(B-R) \approx -0.25(B-R)_0 -(0.09\mu_B +
0.12M_B)$, which implies an anti-correlation of the central color
with the integral color of galaxies, contrary to that inferred from 
direct observational comparisons.

For a sample of 114 spirals, Karachentsev (1991) has found that the ratio
of dynamical mass-to-L$_B$ does not depend on L$_B$ but does anti-correlate
with the $B-$band SB, in agreement with our result (see also Giovanardi
\& Hunt 1988). The anti-correlation in the data of Karachentsev
($\mdynl\propto \Sigma_{B,0}^{-s}$, $s\approx 0.5$) is steeper
than we have found. We expect that the slope $s$ would decrease if 
the velocity-dependent internal extinction correction applied here 
is applied to those data.

Salucci et al. (1991) inferred the ratio \mdynmd, by using
$B-$band data for a sample of HSB galaxies and assuming constant central
SB. They concluded that this ratio decreases roughly as $L_B^{-0.4}$. 
However, as we have mentioned in \S 3.3, a spurious correlation among 
\vdvt\ and luminosity (mass) appears if the sample is restricted in SB. 
The same applies for the \mdynmd\ ratio. Salucci et al. (1991) assumed 
constant SB and their sample was indeed limited to HSB galaxies only. 
Furthermore, we expect that the correlation with  $L_B$ in Salucci et al. will 
diminish if a luminosity-dependent internal extinction coefficient is 
applied to the data.


\begin{figure}[t]
\includegraphics[width=\columnwidth]{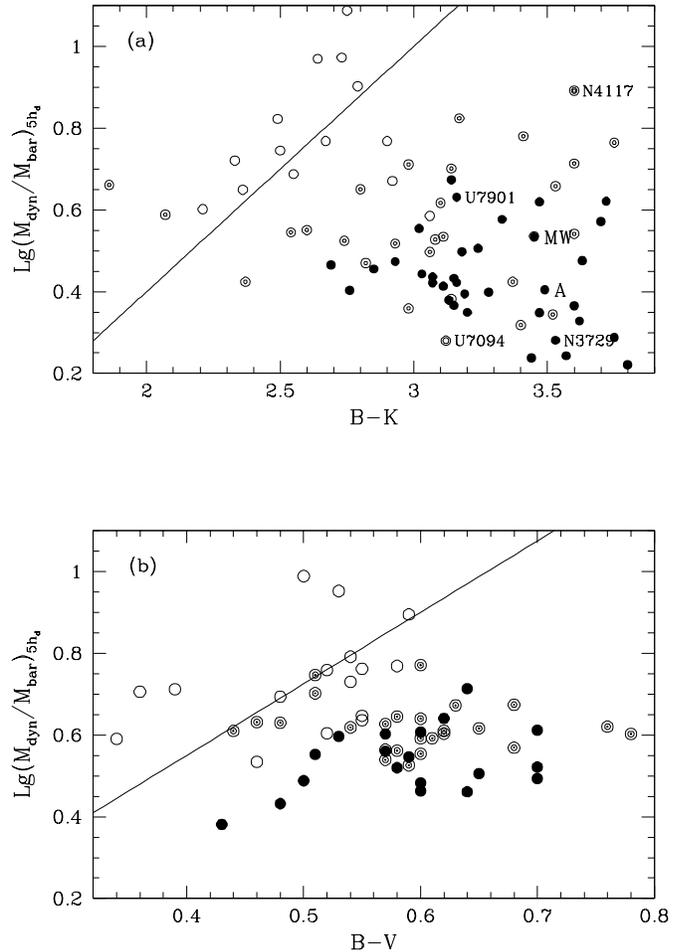}
\caption{{\bf (a).} Approximate
dynamical-to-baryonic mass ratio at 5\hd\ vs. integral
$(B-K)$ color for observational data. No correlation, if any, an
anti-correlation, is seen. However, for limited ranges of surface
density, a correlation appears as this density is lower. The data
in the plot were sorted into three groups according to their disk
central surface densities: open, targeted and solid symbols are
for low, intermedium and high surface densities, respectively (we
use surface densities instead of L$_K$ luminosity in order to have
the same ranges for the model data presented in panel [b]). The
solid line is an eye-fit to the low surface density group. {\bf
(b).} Same as in panel (a) but for the model data and for $(B-V)$
instead of $(B-K)$. See text for explanations of this Figure.}
\end{figure}

 Finally, from the analysis of a sample of optical rotation curves 
from HSB late-type spirals, Persic \& Salucci (1988, 1990) have found that 
\mdynmd\ at 3.2\hd\ decreases with $L_{B}$ as $L_{B}^{-0.4}$. The method
is conceptually different from that used here and it is based on a different 
observational information. Our results (observational and theoretical)
show that $(\mdynmd)_{2.2\hd}$ does not correlate with \md\ or $L_{B}$.   
The main assumption we have done to infer the \mdynmd\ ratio at 2.2\hd\ 
from observations is that $x/f_L(x)$ in eq. (\ref{mdynvt})
remains almost constant with the SB or luminosity. This is indeed
true for disks formed inside CDM halos (or for any halo with a circular
velocity that does not increases significantly after the maximum
of the disk component), where 
$\vdvt \approx (V_{\rm d}/V_{\rm t})_{2.2\hd}$ as seen in Fig. 2d. 
Since our analysis in Sect. 3 of the trends of \vdvt\ on several galaxy 
parameters has shown a reasonable agreement between observations and models,
we may speculate that the galaxy halos do not depart significantly from 
the NFW mass distribution, in particular at intermedium radii. Nevertheless,
more tests are needed in order to see whether $x/f_L(x)$ indeed does not
depend on SB/luminosity. 

A more direct comparison of the results from our observational sample and 
the inferences of Persic and collaborators, can be done through the \vdvt\ 
ratio, a quantity that was calculated from the available data without 
introducing any assumption about $x/f_L(x)$. The analysis presented in \S 3.3 
shows that \vdvt\ for our sample does not 
depend significantly on mass or luminosity, but on SB. In the opposite, 
the mass decomposition of the universal rotation curve implies a strong 
correlation of \vdvt\ with luminosity (see Fig. 6 in Persic et al. 1996),
while the correlation with SB was not explored in that work.
 
The investigation in detail of the differences on the results obtained
with the two different methods is out of the scope of the 
present paper. Our results suggest that the SB should be taken into account
in the kinematical method as an important parameter for the shape of the
rotation curves. On the other hand, we notice that, according to  our
sample data, the \hB/\hd\ (or \hB/\hK) ratio increases on average 
with $L_B$. The results of Persic \& Salucci (1988, 1990) and 
Persic et al. (1996) are based on the determination of the slope of the 
rotation curve at 3.2 scale radii. They used the optical 
(mostly $B-$band) disk scale radii instead of \hK\ or \hd; the latter
are more directly related to the gravitational field of the disk than \hB.  
Since the slope of the rotation curve it is sensible to the radius at which 
is measured, a systematic shift of this radius toward larger values as the 
luminosity increases, will introduce an (extra) anti-correlation of this slope 
with the luminosity, if the the rotation curves are actually more 
decreasing as the galaxy is more luminous or of higher SB. We speculate that, 
by using the more physical \hK\ or \hd\ radii
rather than \hB\ or $h_I$, the anti-correlation of the rotation curve slope 
at 3.2 scale radii with $L_B$ found by the mentioned authors will become 
weaker.

In general, most of the previous works on the dark and luminous contents in 
disk galaxies did not take into account that these contents vary
significantly among galaxies, depending on the disk SB. We have
shown that the ratio of total-to-luminous matter content
within the optical regions of the galaxies decreases as the SB
increases, without any significant (anti)correlation with galaxy scale.

\section{Summarizing discussion}

We have presented a compilation of normal disk galaxies with available 
photometric parameters in both the $B$ and
$K$ bands, rotation curves and/or HI-velocity line-widths,
and HI integral fluxes. A total of 78 galaxies have been selected as
useful for our analysis. We have uniformized the data
(luminosities, disk central SBs and scale radii, maximum rotation
velocity, disk gaseous mass) and used them to further calculate several 
stellar and baryonic galaxy parameters. With this sample we explored 
the ratios of dark to luminous mass (and luminosity) within disk galaxies 
and how these ratios vary with the galaxy properties.  The inferences 
obtained here are alternative to other more direct methods, as 
the rotation curve decomposition and the statistical analysis of rotation curve
shapes. A detailed comparison of our results with those of more direct methods 
is not within the scope of the present paper, but deserves a future analysis. 
In this paper, our observational results were compared with model 
predictions and analyzed further in the light of the behavior of the 
models. We next discuss the main conclusions of this work.

$\bullet$ The \vdvt\ ratio inferred from observations depends mainly
on the disk central surface density, \sigd\ (or \sigK), increasing
approximately as \sigd$^{0.15}$. Only a fraction of HSB 
and ultra-HSB galaxies in our sample fall in the range of 
maximum disk ($\vdvt\approx 0.85\pm 0.10$); as the disk central
surface density decreases, the galaxies become more dark matter 
dominated, with the lowest SB galaxies having \vdvt\ ratios around 
0.5. The MW and M31 galaxies have \vdvt = 0.67 and 0.84, respectively. 
In previous works, by using the analysis of rotation curve shapes,
a ``democratic'' solution to the discussion about maximum or sub-maximum 
disk was found, in the sense that there are galaxies with rotation 
curves highly dominated by the disk component and other ones dominated by 
the halo component (e.g., Persic et al. 1996). The main driving parameter
in these works was the luminosity or mass. Our results show instead that the 
driving parameter is the disk surface brightness or density.

$\bullet$ Although the \vdvt\ ratio depends mainly on \sigd\ (\sigK),
a large scatter is observed in the \vdvt-\sigd\ plot. Interestingly,
we find some trend of decreasing \vdvt\ as the galaxy is
redder, of earlier type and more luminous (massive).
These results hint us about a connexion between the luminous and
morphological galaxy properties to the dark halo properties. The
latter ones are related to the initial (cosmological) conditions.

$\bullet$ To gain a deeper understanding of the observations, we have 
confronted them
with models of disk galaxy formation and evolution in the $\Lambda$CDM
cosmogony (FA00). For a disk-to-total halo mass ratio \fd = 0.05, the models
agree marginally with the observational data in the \vdvt\ vs. \sigd\ plane
(Fig. 3). Models have on average slightly lower \vdvt\ ratios
than observations, the difference becoming larger for the highest values of
\sigd. 
For \fd=0.03, the differences increase slightly. These differences
suggest that real galaxies may have less dark matter in
the center than disks formed within $\Lambda$CDM halos. The observations
in the $\vdvt-\sigd$ plot lie indeed in between models with cuspy NFW halos
and pseudo-isothermal halos with shallow cores, specially at the highest
surface densities, where the differences between both halo models become large.
Nevertheless, the differences
between observations and the $\Lambda$CDM  models are modest, and the
dependences of \vdvt\ on different galaxy properties (in particular \sigd)
for observations and these models agree rather well (e.g., see thick dotted
and solid lines in Fig. 3b). Therefore, only a minor modification to the
inner structure of the $\Lambda$CDM halos seems to be necessary to remedy
the differences with the observations.

A possible mechanism to form soft cores in the CDM halos, without altering
their MAHs and outer structures, is the
gravothermal expansion produced in the core if the CDM particles are allowed to
self-interact among them. Cosmological N-body simulations for self-interacting
CDM particles with a modest cross section inversely
proportional to the particle relative velocity, have shown to be an attractive
alternative to produce soft cores in agreement with observational inferences from
dwarfs to galaxy cluster scales,  keeping at the same time the successful
predictions of the ``standard'' $\Lambda$CDM paradigm (Col\'{\i}n et al. 2002).
The constant expansion of the core could explain why the \vdvt\ ratios of
ultra-HSB galaxies is so high in spite of the strong gravitational drag of the
disk on the halo (see Fig. 2). Models of disk galaxy evolution within the 
self-interacting $\Lambda$CDM halos will show whether self-interaction leads 
to a better agreement with observations than in the case of collisionless CDM 
particles.

$\bullet$ The models show that, in spite that more massive (luminous) systems have
typically less concentrated dark halos than the smaller ones, for a given
\sigd, the former are more dark matter dominated (smaller \vdvt\ ratios)
than the latter. This trend, also seen in the observations, is
easily explained in the light of the models (\S\S 3.1, 3.2.2). On the
other hand, for a given mass, the \vdvt\ ratio of disks in more
concentrated halos is lower than in the less concentrated ones.
More concentrated halos are typically the result of earlier mass assembling,
therefore the disks formed within them are redder because the gas infall rate at
latter epochs is low. Thus, the trend of redder galaxies presenting
smaller \vdvt\ ratios, at least for LSB galaxies, is understandable in the 
models and it might explain the same trend found for the observations. In general,
the scatter of the observational data in the \vdvt-\sigd\ plane
is only slightly larger than for the models (Fig. 3b). Most of this scatter
is explained in the light of the models, and the rest is perhaps due to the
observational uncertainties.

$\bullet$ We also have presented the dynamical-to-baryon mass ratio of models and
observations (Fig. 4). To estimate this ratio at 2.2 and $5 \times \hd$ 
from the kind of observations we have used here, one has to assume that \vt\ 
occurs at 2.2 \hd\ for all the galaxies, or that \vt\ remains constant
up to 5\hd, respectively.  
The \mdynmd\ ratio of observations and models depends mainly
on \sigd\ and does not depend on scale (\hd\ or \md). These trends sharpen
when using (\ms, \sigs) and ($L_K$, \sigK) instead of (\md, \sigd). But when 
passing to ($L_B$, \sigB),  the relation between \mdynl\ and \sigB\ becomes 
shallower and noisier than in the other cases (\mdynms\ vs. \sigs, for example). 
Nevertheless, we find that \mdynl\ still (anti)correlates significantly 
with \sigB, and does not correlate with \hB\ (or $L_B$). 
The \mdyn-to-\md, -\ms, -$L_K$, and -$L_B$ ratios at 5\hd\ for the lowest
SB galaxies are approximately 8, 14, 10.5 and 18, respectively, while for the 
highest SB galaxies are 2, 2, 1.5, and 5, respectively (Figs. 5 and 6). The 
\mdynmd\ ratio for 
the highest SB model galaxies (maximum difference with observations) is 
around 2.4. This means that for the highest SB galaxies, models predict
halos $\sim 1.7$ more massive at 5\hd\ than inferred from 
observations (at 2.2 \hd\ this difference is of $\sim 1.8$).

$\bullet$  The \mdynmd\ ratio does not correlate with $L_B$ (or
\md) as was found previously in works based on the statistical analysis 
of rotation curve shapes, where the determination of the rotation
curve slope at some typical radius (3.2\hB\ in these works) is crucial. 
We have noted that the correlation of \hB/\hd\ (or \hB/\hK) on $L_B$ could 
contribute significantly to an anti-correlation of the slope with 
$L_B$, in the understanding that \hd\ (or \hK) is the physical radius
to be used. We also note that the strong dependence of the \vdvt\ ratio 
on luminosity inferred from the universal rotation curve is not reproduced
by our data analysis. The point is that two different methods, using different 
samples, arrive to different conclusions: we find that the the fractions
of dark and luminous matter within the disks depend mainly on the SB or 
surface density, while in works based on the analysis of the rotation curve 
shapes, the main parameter is the luminosity or mass. In fact
luminosity and SB are correlated, but the determination of which
of them is the dominant is an important issue to understand
galaxy formation and evolution. If the luminosity (or mass) is the
main parameter, then disk formation in $\Lambda$CDM models could 
not be realistic or intermediate astrophysical processes like a
strongly mass-dependent feedback could be relevant. {\it It is important
that future works solve the controversy stated here}.

$\bullet$ The correlations of the mass-to-light ratios with SB in the $B$
and $K$ bands found here imply that the integral color $(B-K)$ correlates with 
the central disk color $(B-K)_0$  (eq.[$\ref{B-K1}$]), in accord to a direct
comparison of both parameters. We have also found that as galaxies are of
less SB, a more significant correlation among the \mdynmd\ ratio and $(B-K)$
appears (Fig. 7). According to the models,  galaxies are redder in part because
they formed in more concentrated dark halos (halos that collapsed earlier
on average). Therefore, for a given SB, \mdynmd\ tends to be larger (more
dominion of dark matter) for redder
galaxies. However, for disks with high SBs, the \mdynmd\ ratio tends to be more
constant than for low SBs (due to the disk gravitational drag over the halo),
in such a way that the correlation with the color minimizes. The similar
features found between models and observations (compare panels a and b of Fig. 7)
suggests that this is a plausible explanation for the observational trend.

$\bullet$ The good agreement found  between models and observations in the 
comparisons presented along the present paper favors the $\Lambda$CDM scenario,
under the assumptions that disk galaxies are formed by gas accretion rather
than mergers,  and that during the collapse of the gas, detailed angular momentum
conservation is obeyed. In the light of the results presented here, the main
difficulty of this scenario is an apparent excess of dark matter within the
disk optical radius. A minor modification to the physics of the formation of the
$\Lambda$CDM halos seems to be enough to overcome this difficulty.

\begin{acknowledgements}

We have made use of the Lyon-Meudon Extragalactic Database (LEDA), 
and the NASA/IPAC Extragalactic Database (NED) which is operated by 
the Jet Propulsion Laboratory, California Institute of Technology, under 
contract with NASA. We thank the referee for useful comments and for
suggesting several improvements to the manuscript. This work was funded 
by CONACyT grant 33776-E to V.A.

\end{acknowledgements}

{\bf Appendix A}

In this Appendix we describe the construction of the disk+halo
composite models. The overall procedure closely resembles that one
presented by Mo et al. (1998).

The disk is assumed to be in centrifugal equilibrium, to be thin, and to 
have an exponential surface density distribution, $\sigd(r)=\sigd exp(-r/\hd)$,
where the central surface density \sigd\ and scale radius \hd\ are related
to the total disk mass by $\md=2\pi\sigd\hd^2$. The circular velocity
is calculated according to eq. (\ref{freeman}). The angular momentum of the
disk, neglecting its gravitational effect, is:
\begin{equation}  \label{jd}
J_d=2\pi \int{V_h(r)\sigd(r) r^2 dr},
\end{equation} 
where V$_h(r)$ is the halo circular velocity. It is assumed that the
disk forms from the baryons trapped within the virialized
halo (with a mass fraction \fd) and that they follow originally the same
mass and angular momentum distribution of the halo. If detailed angular
momentum conservation is obeyed, then $J_d=J$, where $J$ is the halo angular
momentum, and it is related to the dimensionless spin parameter $\lambda$ by:
\begin{equation} \label{lam} 
\lambda = J|E|^{1/2}G^{-1}\mv^{-5/2},
\end{equation}  
where $E$ is the total energy 
of the halo. Now, supplanting these relations into eq. (\ref{jd}), the
disk scale radius \hd\ is calculated; it will be proportional 
to $\lambda$. Therefore, given \mv, \fd, $\lambda$ and the halo density
profile, the two parameters of the exponential disk in centrifugal
equilibrium remain defined. Following, two halo models are analyzed:

{\it Pseudo-isothermal (P-I) sphere.} The density and circular velocity 
profiles of a P-I sphere with core radius r$_c$ and constant central density 
$\rho_0$ are given by:
\begin{eqnarray} \label{p-i}
\rho(r)=\frac{\rho_0}{1+(r/r_c)^2}, \ \ \ \ \ \ \ \ \ \ \ \ \ \ \ \ \ \ \ \ \ \ \ \ \ \ \ \ \ \ \ \ \ \ \ \ \ \ \ \ \ \   \nonumber  \\
V_h^2(r)=\frac{GM_h(r)}{r}=\bigl[4\pi G\rho_0r_c^2\bigl(1-\frac{r_c}{r}arctan(\frac{r}{r_c})\bigr)\bigr] 
\end{eqnarray} 
The asymptotic velocity is V$_{\rm asym}=(4\pi G \rho_0 r_c^2)^{0.5}$.
According to the cosmological spherical collapse model, at the epoch
of virialization of the last shell of an overdense homogeneous sphere, 
its radius (virial radius, r$_v$) is such that $<\rho(<r_v)>=\Delta_c \rho_u$,
where $<\rho(<r)>$ is the average density of the sphere at $r$, 
$\Delta_c$ is a constant, and $\rho_u$ is the matter density of the universe 
at that epoch. 
For the present epoch and for the $\Lambda$CDM cosmology used here, 
$\Delta_c=334$ (Eke, Navarro \& Frenk 1998). The mass contained within
r$_v$ is the virial (halo) mass, \mv. Taking into account that r$_v$/r$_c$$>>1$,
it is easy to find that V$_{\rm asym}\approx (G\mv/r_v)^{1/2}\approx
0.014(\mv/\msun)^{1/3}\kms$, and that $E\approx -(\mv V_{\rm asym}^2)/2$.  
At radii larger than r$_c$, V$_h$ rapidly tends to V$_{\rm asym}$=const, and 
the integral in eq. (\ref{jd}) can be approximated by 
$J_d\approx 2\md \hd$V$_{\rm asym}$. Now, from $J_d=J$ and using eq. 
(\ref{lam}), \hd\ can be calculated, \hd $\propto \lambda \mv^{1/3} \fd^{-1}$,
and $\sigd \propto \mv^{1/3}\fd^3 \lambda^{-2}$. 

Once \mv, \fd, and $\lambda$ are defined, we calculate the disk parameters 
\hd\ and \sigd\ and use them in eq. (\ref{freeman}) to get the disk velocity
component. The halo velocity component is calculated according to eq. (\ref{p-i}).
The core radius r$_c$ is  taken from empirical inferences found
by Firmani et al. (2001) from dwarf and LSB galaxies:  r$_c'\approx 
5.5($V$_{\rm asym}$/100\kms)$^{0.95}$, where we have to take into account that 
the core radius of the non-singular isothermal model used by them, r$_c'$, 
is 1.45 times the core radius of the pseudo-isothermal model used here, 
r$_c$= r$_c'$/1.45. 

{\it NFW cosmological halo model.} The typical density and circular velocity 
profiles of the CDM halos are given by:
\begin{eqnarray} \label{nfw}
\rho(r)=\frac{\rho_s}{(r/r_s)(1+r/r_s)^2}, \ \ \ \ \ \ \ \ \ \ \ \ \ \ \ \ \ \ \ \ \ \ \ \ \ \ \ \ \ \   \nonumber \\
V_h^2(r)=\frac{GM_h(r)}{r}=\frac{4\pi G \rho_s r_s^3\bigl[ln(1+\frac{r}{r_s})-
\frac{1}{1+r_s/r}\bigr]}{r},
\end{eqnarray}
where the parameters $\rho_s$ and r$_s$ are related. Defining again as above
the virial radius r$_v$ and introducing the concentration parameter $c=r_v/r_s$,
one obtains that $\rho_s= (\Delta_c \rho_u/3)c^3 g(c)$, where 
$g(c)=1/(ln(1+c)-c/(1+c))$. Now, the halo circular velocity can be expressed
in terms of the velocity at the virial radius (V$_v$=(G\mv /r$_v$)$^{1/2}$):
(V$_h/$V$_v$)$^2=(1/w)(ln(1+cw)-cw/(1+cw))g(c)$, where $w=r/$r$_v$. As shown above,
from the spherical collapse model one obtains that V$_v\propto \mv^{1/3}$.
Therefore, given \mv\ and c, the NFW velocity profile remains defined.
In order to calculate the disk parameters, the total halo energy is necessary:
$E=-G\mv^2 f(c)/(2$r$_v)$, where $f(c)$ is a function of the concentration
parameter (see Mo et al. 1998) and r$_v\propto $V$_v$. Cosmological numerical
simulations show that c slightly decreases with \mv\ and has a significant
scatter ($\Delta$log$c_{\rm NFW}\approx 0.18$, Bullock et al. 2001).

{\it Gravitational contraction of the NFW halo.} The formation
of a dense disk inside the cosmological dark halo induces a gravitational 
contraction of the halo in the inner regions. Under the assumption of 
radial adiabatic invariance, one may calculate the mean radius $r$ where 
ends a particle with an initial mean radius $r_i$: $GM_f(r)r=GM(r_i)r_i$,
where $M_f(r)=M_d(r)+M(r_i)(1-\fd)$ is the final total mass distribution
and $M(r_i)$ is the initial halo (NFW) mass distribution. To calculate the 
disk mass distribution $M_d(r)$, \hd\ should be found as described above.
But to calculate \hd\ the total circular velocity V$_t^2(r)=$V$_d^2(r)+$V$_h^2(r)$ 
should be given, where V$_h^2(r)=G[M_f(r)-M_d(r)]/r$. Given \mv, \fd, $\lambda$
and c, an iterative procedure yields \hd\ and V$_t(r)$ after the halo
contraction. This procedure is explained in Mo et. al (1998) and we refer
the reader to that paper for more details.

{\bf Appendix B.}

Here we present the main ingredients of the self-consistent evolutionary
models to be studied in this paper. 
The disk is built up within a growing $\Lambda$CDM halo. A special extended
Press-Schechter approach is used to generate the statistical mass aggregation
histories (MAHs) of the halos from the primordial density fluctuation field, 
and a generalized secondary infall model is applied to calculate the 
virialization of the accreting mass shells. The evolution and structure of 
the $\Lambda$CDM halos calculated this way agree well with results
from cosmological N-body simulations (Avila-Reese et al. 1999); halos
assembled through early active MAHs end more concentrated on average
than halos with extended MAHs. The mass
shells are assumed to have aligned rotation axis with specific angular 
momentum given by $j_{sh}(t_v)=dJ(t_v)/dM_v(t_v)$,  where
$J=\lambda GM_v^{5/2}/\left| E\right| ^{1/2}$, $J$, M$_v$ and $E$ are
the total angular momentum, mass and energy of the halo at the shell
virialization time $t_v$. The spin parameter, $\lambda$, is assumed to
be constant in time. As the result of the assembling of these
mass shells, a present day halo ends with an angular momentum distribution
close to the (universal) distribution measured by Bullock et al. (2001) 
in N-body simulations. A fraction \fd\ of the mass of each shell is assumed
to cool down and form a disk layer in a dynamical time. The radial mass 
distribution of the layer is calculated by equating its specific angular
momentum to that of its final circular orbit in centrifugal equilibrium
(detailed angular momentum conservation is assumed). The supperposition
of these layers form the disk. The gravitational
interaction of disk and halo is calculated using the adiabatic invariance
formalism. The local SF is triggered by the Toomre gas gravitational
instability criterion and self-regulated by a vertical disk balance
between the energy input due to SNe and the turbulent energy dissipation
in the ISM. The SF efficiency depends on the gas surface
density determined mainly by $\lambda$, and on the gas accretion
rate determined by the cosmological MAH.
Finally, we calculate the formation of a secular bulge by using the Toomre
instability criterion for the stellar disk.

\clearpage

\begin{table*}
\caption[]{Raw and corrected photometric and dynamical parameters of the disk
galaxy sample}
\label{tabgeneraldata}
\begin{tabular}{lccccccccc} 
\hline\noalign{\smallskip}
Name & D$^{\mathrm{a}}$ & $m_{B}^{\mathrm{b}}$ & $m_{K}^{\mathrm{b}}$ &
$\mu_{0,B}^{\mathrm{c}}$&$\mu_{0,K}^{\mathrm{c}}$ & $(B-K)^{\mathrm{d}}$ & 
$h_{B}^{\mathrm{e}}$ & W$_{20,c}^{\mathrm{f}}$ & Ref.$^{\mathrm{k}}$ \\
T & $i(^{o})$ & Log$L_{B,c}^{\mathrm{g}}$ & Lg$L_{K,c}^{\mathrm{g}}$ & 
Lg$\Sigma_{0,Bc}^{\mathrm{h}}$ & Lg$\Sigma_{0,Kc}^{\mathrm{h}}$ & 
$(B-K)_{0}^{\mathrm{i}}$ & $h_{K}^{\mathrm{e}}$ & LgM$_{HI}^{\mathrm{j}}$ & 
\\
\hline\noalign{\smallskip}
U89&67.6&12.63&8.85&22.07&17.46&3.40&9.34&516.4&"1,6"\\
1&49.3&10.99&11.52&1.97&2.83&4.22&5.70&9.96&\\ \hline
U242&64.6&13.92&10.30&21.26&17.28&3.24&4.29&407.3&"1,6"\\
6&40.6&10.43&10.90&2.35&2.96&3.59&3.48&9.75&\\ \hline
U334&68.7&15.24&12.07&23.36&20.32&2.92&7.16&187.1&"1,6"\\
9&41.6&9.90&10.24&1.40&1.74&2.91&6.02&9.80&\\ \hline
U438&66.9&12.95&9.10&20.45&16.23&3.57&4.57&547.4&"1,6"\\
5&39.7&10.81&11.41&2.63&3.38&3.96&3.73&9.91&\\ \hline
U490&66.9&13.48&9.46&21.47&17.16&3.62&6.00&471.7&"1,6"\\
5&46.0&10.65&11.27&2.24&2.97&3.91&4.83&10.18&\\ \hline
U628&79.8&15.58&12.52&22.86&20.39&2.73&5.73&260.7&"1,6"\\
9&52.4&9.93&10.20&1.49&1.63&2.40&7.20&9.82&\\ \hline
U1305&39.2&12.75&8.65&22.02&17.61&3.70&6.44&513.9&"1,6"\\
4&36.8&10.47&11.12&2.09&2.86&4.01&5.03&9.61&\\ \hline
U1577&77.9&13.79&9.90&22.44&18.26&3.60&8.34&423.4&"1,6"\\
3&38.7&10.61&11.22&1.78&2.58&4.06&6.64&10.11&\\ \hline
U1719&120.3&14.08&10.09&22.45&17.73&3.53&12.65&690.4&"1,6"\\
3&44.2&10.95&11.54&1.86&2.75&4.29&8.34&10.13&\\ \hline
U2064&62.2&14.20&10.18&22.28&18.01&3.37&6.06&328.1&"1,6"\\
4&46.8&10.41&10.92&2.06&2.64&3.53&5.31&9.82&\\ \hline
U2081&36.8&14.50&11.55&22.31&19.44&2.67&3.40&216.5&"1,6"\\
6&54.7&9.66&9.90&1.69&1.99&2.82&3.46&9.33&\\ \hline
U4308&52.1&13.17&9.83&21.34&17.66&3.06&5.05&370.7&"1,6"\\
5&40.6&10.50&10.89&2.29&2.81&3.38&4.14&9.94&\\ \hline
U4368&56.7&13.57&10.21&21.52&17.81&3.11&4.84&369.8&"1,6"\\
6&37.8&10.40&10.82&2.22&2.77&3.45&4.04&10.04&\\ \hline
U4375&30.3&13.28&9.53&21.31&17.21&3.33&2.97&355.5&"1,6"\\
5&50.1&10.04&10.55&2.31&2.93&3.62&2.49&9.26&\\ \hline
U4422&63.2&12.85&8.83&22.04&18.38&3.75&9.51&606.5&"1,6"\\
5&35.7&10.79&11.46&1.94&2.55&3.59&8.28&10.11&\\ \hline
U5103&54.8&12.77&9.22&20.50&16.56&3.15&4.72&465.3&"1,6"\\
3&54.0&10.76&11.19&2.58&3.15&3.50&3.85&9.98&\\ \hline
U5303&20.9&12.24&8.67&21.32&17.89&3.14&3.73&330.0&"1,6"\\
5&55.5&10.15&10.57&2.29&2.61&2.89&3.83&9.38&\\ \hline
U5510&19.8&12.36&9.24&20.66&17.37&2.93&1.96&287.5&"1,6"\\
5&36.8&9.94&10.28&2.56&2.96&3.05&1.77&9.42&\\ \hline
U5554&18.7&12.88&9.10&20.98&16.94&3.47&1.73&355.3&"1,6"\\
1&47.6&9.74&10.30&2.43&3.06&3.65&1.52&8.69&\\ \hline
U5633&20.5&14.24&11.28&23.13&20.01&2.64&2.77&198.8&"1,6"\\
8&52.4&9.27&9.50&1.43&1.80&2.99&2.93&9.19&\\ \hline
U6028&16.9&13.14&10.19&20.49&17.27&2.60&1.25&233.6&"1,6"\\
3&56.2&9.56&9.77&2.62&2.85&2.66&1.06&9.14&\\ \hline
U6453&17.8&12.45&9.18&20.91&17.16&3.02&1.84&331.0&"1,6"\\
4&42.5&9.84&10.22&2.46&3.01&3.44&1.41&9.47&\\ \hline
U6460&17.7&11.87&8.62&20.66&17.15&3.03&2.37&273.9&"1,6"\\
4&42.5&10.05&10.44&2.56&3.01&3.20&2.07&9.02&\\ \hline
U6746&102.6&13.35&9.39&21.38&17.30&3.72&8.80&798.1&"1,6"\\
0&41.6&11.01&11.67&2.21&2.93&3.89&7.96&9.94&\\ \hline
U7169&32.4&12.62&9.63&20.11&16.55&2.76&2.08&347.3&"1,6"\\
5&36.8&10.28&10.56&2.79&3.28&3.30&1.57&9.65&\\ \hline
U7315&13.7&12.55&8.62&19.99&16.03&3.60&0.98&313.5&"1,6"\\
4&49.3&9.60&10.22&2.83&3.41&3.53&0.97&8.39&\\ \hline
U7523&14.7&11.80&8.08&21.49&17.70&3.52&2.57&264.1&"1,6"\\
3&36.8&9.91&10.49&2.25&2.83&3.52&2.32&8.55&\\ \hline
U7901&13.0&11.46&7.90&20.07&16.06&3.16&1.58&477.9&"1,6"\\
5&49.3&10.03&10.46&2.79&3.40&3.59&1.24&9.39&\\ \hline
\hline\noalign{\smallskip}
\end{tabular}
\end{table*}

\setcounter{table}{0}

\begin{table*}
  \caption[]{Continued.}

\begin{tabular}{lccccccccc}

\hline\noalign{\smallskip}
Name & D$^{\mathrm{a}}$ & $m_{B}^{\mathrm{b}}$ & $m_{K}^{\mathrm{b}}$ &
$\mu_{0,B}^{\mathrm{c}}$&$\mu_{0,K}^{\mathrm{c}}$ & $(B-K)^{\mathrm{d}}$ & 
$h_{B}^{\mathrm{e}}$ & W$_{20,c}^{\mathrm{f}}$ & Ref.$^{\mathrm{k}}$ \\
T & $i(^{o})$ & Log$L_{B,c}^{\mathrm{g}}$ & Lg$L_{K,c}^{\mathrm{g}}$ & 
Lg$\Sigma_{0,Bc}^{\mathrm{h}}$ & Lg$\Sigma_{0,Kc}^{\mathrm{h}}$ & 
$(B-K)_{0}^{\mathrm{i}}$ & $h_{K}^{\mathrm{e}}$ & LgM$_{HI}^{\mathrm{j}}$ & 
\\
\hline\noalign{\smallskip}
U8279&39.7&13.07&9.74&20.52&16.90&3.16&2.62&369.8&"1,6"\\
4&40.6&10.25&10.69&2.58&3.12&3.41&2.21&9.60&\\ \hline
U8865&36.9&12.62&8.98&21.89&18.27&3.41&5.73&422.4&"1,6"\\
2&41.6&10.39&10.93&1.96&2.56&3.59&5.24&9.62&\\ \hline
U9481&56.7&13.31&10.00&21.22&18.03&3.14&4.84&322.2&"1,6"\\
4&43.4&10.47&10.90&2.29&2.64&2.95&4.81&10.08&\\ \hline
U9926&30.3&12.17&8.45&20.13&16.45&3.28&2.63&461.4&"1,6"\\
5&46.0&10.50&10.98&2.81&3.27&3.22&2.44&9.46&\\ \hline
U9943&30.1&12.30&8.67&20.40&16.60&3.19&2.91&401.9&"1,6"\\
5&51.7&10.44&10.89&2.67&3.16&3.31&2.56&9.60&\\ \hline
U10083&28.7&12.37&9.19&21.51&17.60&2.98&3.24&270.5&"1,6"\\
2&35.7&10.26&10.63&2.24&2.87&3.64&2.63&9.65&\\ \hline
U10445&16.8&13.10&10.81&21.76&19.00&2.07&1.65&216.2&"1,6"\\
6&40.6&9.51&9.51&2.05&2.29&2.66&1.59&9.25&\\ \hline
U11628&61.6&12.82&8.54&22.27&17.19&3.63&11.60&659.9&"1,6"\\
2&46.8&10.95&11.58&2.01&2.96&4.46&6.55&10.23&\\ \hline
U11872&18.1&12.03&8.11&20.46&15.71&3.44&1.61&420.4&"1,6"\\
3&43.4&10.12&10.67&2.72&3.59&4.23&1.11&9.31&\\ \hline
U12343&35.2&11.87&7.81&21.95&17.67&3.45&8.65&544.2&"1,6"\\
4&40.6&10.82&11.37&2.18&2.82&3.67&6.77&10.04&\\ \hline
U12511&52.0&13.81&10.58&22.46&18.35&2.98&6.15&434.6&"1,6"\\
6&35.7&10.23&10.59&1.78&2.56&4.02&3.33&10.32&\\ \hline
U12614&51.8&12.64&9.12&20.94&17.24&3.15&5.43&420.1&"1,6"\\
5&44.2&10.75&11.18&2.46&2.956&3.32&5.08&9.92&\\ \hline
U12638&82.0&13.95&9.99&22.17&18.27&3.60&8.51&469.8&"1,6"\\
5&36.8&10.62&11.23&1.92&2.59&3.74&7.88&10.03&\\ \hline
U12654&59.5&13.65&10.28&21.76&17.98&3.08&5.65&354.2&"1,6"\\
4&35.7&10.43&10.83&2.16&2.71&3.45&4.38&9.85&\\ \hline
U6399&14.6&14.33&11.09&21.83&18.72&2.79&1.69&172.0&"2,6"\\
9&75.0&9.00&9.29&1.56&1.96&3.07&1.86&9.03&\\ \hline
U6446&12.5&13.52&11.50&22.61&19.31&1.86&2.28&174.0&"2,6"\\
7&51.0&9.08&8.99&1.59&2.08&3.30&1.49&9.19&\\ \hline
N3726&15.2&11.00&7.96&21.07&17.19&2.74&4.30&331.0&"2,6"\\
5&53.0&10.32&10.60&2.37&2.93&3.48&2.79&9.82&\\ \hline
N3769&13.5&12.80&9.10&19.93&16.34&3.13&1.26&256.0&"2,6"\\
3&70.0&9.64&10.06&2.76&2.98&2.62&1.30&9.37&\\ \hline
N3877&15.9&11.91&7.75&19.72&15.50&3.20&2.60&335.0&"2,6"\\
5&76.0&10.31&10.76&2.81&3.16&2.93&2.41&9.21&\\ \hline
N3893&17.0&11.20&7.84&19.88&16.71&3.07&2.13&382.0&"2,6"\\
5&49.0&10.34&10.74&2.86&3.16&2.82&2.23&9.75&\\ \hline
N3917&17.1&12.66&9.08&20.59&17.12&2.80&2.69&276.0&"2,6"\\
6&79.0&9.98&10.28&1.99&2.55&3.46&2.84&9.26&\\ \hline
N3949&14.5&11.55&8.43&19.54&16.55&2.85&1.31&321.0&"2,6"\\
4&55.0&10.06&10.38&2.99&3.20&2.60&1.35&9.31&\\ \hline
N3953&18.4&11.03&7.03&20.43&16.47&3.47&3.85&446.0&"2,6"\\
4&62.0&10.59&11.16&2.62&3.14&3.37&3.79&9.41&\\ \hline
N3972&15.5&13.09&9.39&20.23&16.50&3.10&1.71&264.0&"2,6"\\
4&77.0&9.66&10.07&2.62&2.87&2.71&1.62&8.95&\\ \hline
U6917&16.2&13.15&10.30&22.27&19.17&2.49&2.93&224.0&"2,6"\\
9&56.0&9.526&9.69&1.68&2.06&3.02&2.55&9.26&\\ \hline
U6923&18.6&13.91&11.04&21.42&18.34&2.54&1.62&160.0&"2,6"\\
10&65.0&9.33&9.51&1.91&2.28&3.01&1.57&8.92&\\ \hline
N3992&18.3&10.86&7.23&20.29&16.82&3.14&4.05&547.0&"2,6"\\
4&56.0&10.65&11.07&2.69&3.05&2.97&4.11&9.76&\\ \hline
U6983&18.8&13.10&10.52&22.59&19.41&2.36&3.60&221.0&"2,6"\\
6&49.0&9.63&9.74&1.64&2.07&3.15&2.67&9.51&\\ \hline
\hline\noalign{\smallskip}
\end{tabular}
\end{table*}

\setcounter{table}{0}

\begin{table*}
  \caption[]{Continued.}

\begin{tabular}{lccccccccc}

\hline\noalign{\smallskip}
Name & D$^{\mathrm{a}}$ & $m_{B}^{\mathrm{b}}$ & $m_{K}^{\mathrm{b}}$ &
$\mu_{0,B}^{\mathrm{c}}$&$\mu_{0,K}^{\mathrm{c}}$ & $(B-K)^{\mathrm{d}}$ & 
$h_{B}^{\mathrm{e}}$ & W$_{20,c}^{\mathrm{f}}$ & Ref.$^{\mathrm{k}}$ \\
T & $i(^{o})$ & Log$L_{B,c}^{\mathrm{g}}$ & Lg$L_{K,c}^{\mathrm{g}}$ & 
Lg$\Sigma_{0,Bc}^{\mathrm{h}}$ & Lg$\Sigma_{0,Kc}^{\mathrm{h}}$ & 
$(B-K)_{0}^{\mathrm{i}}$ & $h_{K}^{\mathrm{e}}$ & LgM$_{HI}^{\mathrm{j}}$ & 
\\
\hline\noalign{\smallskip}
N4051&13.1&10.98&7.86&20.78&16.90&2.93&2.94&308.0&"2,6"\\
4&49.0&10.17&10.51&2.48&3.09&3.58&2.10&9.26&\\ \hline
N4088&14.1&11.23&7.46&19.77&16.19&3.11&2.33&362.0&"2,6"\\
5&69.0&10.33&10.75&2.84&3.11&2.74&2.54&9.72&\\ \hline
N4100&18.6&11.91&8.02&19.82&15.77&3.07&2.81&386.0&"2,6"\\
4&73.0&10.39&10.79&2.80&3.18&3.03&2.54&9.63&\\ \hline
N4102&15.4&12.04&7.86&19.28&15.82&3.80&1.30&393.0&"2,6"\\
3&56.0&9.97&10.66&3.08&3.45&3.00&1.34&8.75&\\ \hline
N3718&17.5&11.28&7.47&21.88&17.52&3.17&6.83&476.0&"2,6"\\
1&69.0&10.50&10.94&2.00&2.63&3.66&4.79&9.86&\\ \hline
N3729&18.0&12.31&8.60&20.22&16.44&3.53&1.67&296.0&"2,6"\\
1&49.0&9.89&10.48&2.70&3.27&3.50&1.62&9.14&\\ \hline
U6773&16.4&14.42&11.23&21.67&18.79&3.06&1.19&112.0&"2,6"\\
10&58.0&8.92&9.32&1.90&2.21&2.84&1.33&8.55&\\ \hline
U6818&14.6&14.43&11.70&21.62&18.68&2.33&1.49&151.0&"2,6"\\
7&75.0&8.94&9.05&1.65&1.97&2.88&1.41&8.90&\\ \hline
N3985&16.7&13.25&10.19&20.04&17.06&2.82&0.87&180.0&"2,6"\\
9&51.0&9.46&9.75&2.79&2.97&2.52&1.02&8.84&\\ \hline
U7089&14.0&13.73&11.11&21.51&18.73&2.21&2.24&138.0&"2,6"\\
8&80.0&9.19&9.24&1.51&1.78&2.74&2.32&9.23&\\ \hline
U7094&14.1&14.74&11.58&21.99&18.56&3.12&1.31&76.0&"2,6"\\
8&70.0&8.62&9.04&1.70&2.24&3.40&1.10&8.75&\\ \hline
N4117&16.3&14.05&9.98&21.16&17.80&3.60&1.04&285.0&"2,6"\\
-2&68.0&9.22&9.83&2.01&2.52&3.33&1.18&8.97&\\ \hline
N4138&15.9&12.27&8.19&20.05&15.98&3.75&1.25&374.0&"2,6"\\
-1&53.0&9.85&10.52&2.77&3.40&3.65&1.20&8.71&\\ \hline
N4218&13.5&13.69&10.83&20.35&16.74&2.69&0.63&150.0&"2,6"\\
1&53.0&9.07&9.31&2.65&3.08&3.14&0.51&8.49&\\ \hline
N4220&16.4&12.34&8.36&20.08&15.27&3.18&1.96&399.0&"2,6"\\
0&78.0&10.09&10.54&2.69&3.40&3.84&1.38&8.99&\\ \hline
U128&67.8&15.16&12.10&23.55&20.30&2.75&7.98&336.6&"3,4"\\
8&36.0&9.94&10.22&1.37&1.78&3.09&6.90&9.99&\\ \hline
F563-V2&69.6&16.25&13.80&22.16&19.29&2.37&2.63&152.5&"3,4"\\
10&45.1&9.43&9.55&1.79&2.12&2.90&2.16&9.46&\\ \hline
F568-3&93.5&16.12&13.11&22.33&19.16&2.90&4.03&241.1&"3,4"\\
7&39.7&9.76&10.09&1.75&2.20&3.21&3.62&9.60&\\ \hline
F574-1&106.0&16.67&13.80&23.00&19.66&2.55&5.91&196.5&"3,4"\\
7&67.9&9.74&9.94&1.20&1.73&3.39&4.26&9.61&\\ \hline
F583-1&38.3&16.40&13.50&23.05&19.80&2.50&2.30&173.3&"3,4"\\
9&65.7&8.99&9.16&1.31&1.73&3.12&1.49&9.40&\\ \hline
MilkyWay&-&-&-&-&-&3.45&5.00&440.0&"9,10"\\
4&-&10.28&10.83&1.98&3.00&4.62&3.00&9.60&\\ \hline
Andromeda&0.7&4.36&-&-&15.67&3.49&6.40&525.4&"7,8,9"\\
3&75.2&10.70&11.27&2.29&3.24&4.46&4.10&2.72&\\ \hline \hline
F563-1&49.0&15.60&-&23.53&-&-&4.67&118.6&"4"\\
9&52.4&9.38&-&1.20&-&-&-&-&\\ \hline
F563-V1&54.8&16.93&-&23.54&-&-&2.58&77.0&"4"\\
10&42.5&8.91&-&1.26&-&-&-&-&\\ \hline
F565-V2&51.9&17.58&-&23.98&-&-&2.90&97.1&"4"\\
10&65.0&8.62&-&0.87&-&-&-&-&\\ \hline
F568-V1&87.0&16.67&-&23.01&-&-&3.46&273.9&"4"\\
8&35.7&9.43&-&1.47&-&-&-&-&\\ \hline
F571-V1&85.6&17.40&-&23.78&-&-&3.46&143.8&"4"\\
8&35.7&9.10&-&1.16&-&-&-&-&\\ \hline
ESO-LV1150280&91.3&14.32&-&21.19&-&-&4.86&212.0&"5,6,11"\\
4&53.2&10.52&-&2.14&-&-&-&-&\\ \hline
\hline\noalign{\smallskip}
\end{tabular}
\end{table*}

\setcounter{table}{0}

\begin{table*}
  \caption[]{Continued.}

\begin{tabular}{lccccccccc}

\hline\noalign{\smallskip}
Name & D$^{\mathrm{a}}$ & $m_{B}^{\mathrm{b}}$ & $m_{K}^{\mathrm{b}}$ &
$\mu_{0,B}^{\mathrm{c}}$&$\mu_{0,K}^{\mathrm{c}}$ & $(B-K)^{\mathrm{d}}$ & 
$h_{B}^{\mathrm{e}}$ & W$_{20,c}^{\mathrm{f}}$ & Ref.$^{\mathrm{k}}$ \\
T & $i(^{o})$ & Log$L_{B,c}^{\mathrm{g}}$ & Lg$L_{K,c}^{\mathrm{g}}$ & 
Lg$\Sigma_{0,Bc}^{\mathrm{h}}$ & Lg$\Sigma_{0,Kc}^{\mathrm{h}}$ & 
$(B-K)_{0}^{\mathrm{i}}$ & $h_{K}^{\mathrm{e}}$ & LgM$_{HI}^{\mathrm{j}}$ & 
\\
\hline\noalign{\smallskip}
ESO-LV1530170&90.9&13.99&-&23.07&-&-&12.92&246.9&"5,6,11"\\
5&45.9&10.62&-&1.43&-&-&-&-&\\ \hline
ESO-LV2520100&140.0&14.20&-&22.69&-&-&13.09&621.1&"5,6,11"\\
2&41.6&10.89&-&1.55&-&-&-&-&\\ \hline
ESO-LV3500110&221.1&14.49&-&22.72&-&-&15.73&404.0&"5,6,11"\\
1&40.6&11.11&-&1.48&-&-&-&-&\\ \hline
ESO-LV5520190&173.6&15.07&-&22.75&-&-&17.24&545.7&"5,6,11"\\
2&53.2&10.84&-&1.46&-&-&-&-&\\ \hline
ESO-LV1220040&139.4&15.47&-&23.96&-&-&26.91&112.5&"5,6,11"\\
3&52.4&10.45&-&1.11&-&-&-&-&\\ \hline
ESO-LV3740090&39.9&15.45&-&23.40&-&-&3.23&112.2&"5,6,11"\\
1&36.8&9.38&-&1.51&-&-&-&-&\\
\hline\noalign{\smallskip}
\end{tabular}

\begin{list}{}{}
\item[$^{\mathrm{a}}$] Luminous distance in Mpc.
\item[$^{\mathrm{b}}$] Raw uncorrected apparent magnitude.
\item[$^{\mathrm{c}}$] Raw uncorrected disk central SB (magarcsec$^{-2}$).
\item[$^{\mathrm{d}}$] Corrected integral $(B-K)$ color.
\item[$^{\mathrm{e}}$] Disk scale radius in kpc.
\item[$^{\mathrm{f}}$] Corrected HI equivalent line-width at 20\% 
level of intensity (km/s).
\item[$^{\mathrm{g}}$] Corrected total luminosity (in $L_{B\odot}$ and 
$L_{K\odot}$, respectively).  
\item[$^{\mathrm{h}}$] Corrected disk central SB (in  $L_{B\odot}$pc$^{-2}$
and $L_{K\odot}$pc$^{-2}$, respectively).
\item[$^{\mathrm{i}}$] Corrected disk central $(B-K)$ color.
\item[$^{\mathrm{j}}$] HI mass in \msun.
\item[$^{\mathrm{k}}$] References: 1.- Verheijen (1997); Verheijen \& 
Sancisi (2001).
2.- de Jong (1996a, 1996b). 3.- Bell et al. (2000). 4.- de Blok et 
al.(1995,1996,2001). 5.- Beijersbergen et al. (1999). 6.- LEDA.
7.- Cox 2000.
8.- Hiromoto et al. (1983). 9.- Gilmore et al. (1990). 10.- Kent et al. (1991).
11.- Graham (2002).

\end{list}
\end{table*}

\clearpage

\begin{table*}
\caption{Inferred stellar, gaseous, and baryonic parameters}
\label{tabgeneraldata2}
\begin{tabular}{cccccccccccc} 
\hline\noalign{\smallskip}

Name  &Lg\ms$^{\mathrm{a}}$ & Lg\sigs$^{\mathrm{b}}$ & 
Lg\mg$^{\mathrm{c}}$ & Lg$\Sigma_{g}^{\mathrm{d}}$ & f$_{g}^{\mathrm{e}}$ & 
Lg\md$^{\mathrm{f}}$ & Lg\sigd$^{\mathrm{g}}$ & \hd$^{\mathrm{h}}$ & 
\vdvt$^{\mathrm{i}}$ & \mdynmd$^{\mathrm{j}}$ &  \mdynl$^{\mathrm{k}}$ 
\\
\hline\noalign{\smallskip}
U89&11.33&2.63&10.11&0.84&0.06&11.35&2.64&6.06&0.64&2.08&4.79\\
U242&10.67&2.74&10.06&1.22&0.20&10.77&2.75&3.87&0.74&3.21&6.94\\
U334&9.99&1.49&9.94&0.63&0.47&10.26&1.54&8.46&0.59&4.69&10.92\\
U438&11.25&3.22&10.31&1.41&0.10&11.29&3.23&3.95&0.97&1.75&5.36\\
U490&11.12&2.83&10.57&1.45&0.22&11.23&2.84&5.56&0.85&2.13&8.08\\
U628&10.04&1.47&9.96&0.49&0.45&10.30&1.51&9.54&0.44&9.40&22.08\\
U1305&10.99&2.73&10.10&0.94&0.11&11.04&2.74&5.34&0.68&3.73&13.98\\
U1577&10.92&2.28&10.69&1.29&0.37&11.12&2.32&8.79&0.65&3.48&11.31\\
U1719&11.37&2.58&10.71&1.12&0.18&11.46&2.59&9.38&0.57&4.55&14.50\\
U2064&10.72&2.44&10.30&1.10&0.28&10.86&2.46&6.17&0.83&2.66&7.59\\
U2081&9.78&1.87&9.64&0.81&0.42&10.01&1.90&4.43&0.56&5.87&13.14\\
U4308&10.63&2.55&10.33&1.35&0.34&10.81&2.57&5.03&0.76&3.14&6.42\\
U4368&10.56&2.51&10.35&1.38&0.38&10.77&2.54&5.03&0.73&3.43&8.02\\
U4375&10.34&2.72&9.66&1.12&0.17&10.42&2.73&2.71&0.70&3.78&9.04\\
U4422&11.16&2.25&10.51&0.92&0.18&11.25&2.27&9.65&0.45&5.82&16.67\\
U5103&10.95&2.90&10.55&1.63&0.29&11.09&2.92&4.57&0.86&2.33&4.97\\
U5303&10.32&2.36&9.78&0.86&0.22&10.43&2.38&4.26&0.62&5.03&9.67\\
U5510&9.99&2.66&9.82&1.57&0.40&10.21&2.70&2.24&0.75&3.30&6.22\\
U5554&10.12&2.88&8.83&0.72&0.05&10.14&2.88&1.56&0.63&4.17&10.48\\
U5633&9.39&1.68&9.36&0.67&0.48&9.67&1.72&3.83&0.46&9.33&23.44\\
U6028&9.41&2.49&9.71&1.91&0.67&9.89&2.59&1.73&0.72&3.56&7.63\\
U6453&9.94&2.73&9.96&1.91&0.51&10.25&2.79&2.01&0.69&3.59&9.36\\
U6460&10.16&2.74&9.51&1.12&0.18&10.25&2.75&2.26&0.84&2.78&4.40\\
U6746&11.54&2.81&10.08&0.53&0.03&11.56&2.81&8.12&0.59&4.18&14.81\\
U7169&10.23&2.95&10.05&1.90&0.40&10.45&2.99&2.02&0.83&2.53&3.71\\
U7315&10.06&3.26&8.88&1.15&0.06&10.09&3.26&1.00&0.88&2.32&7.13\\
U7523&10.32&2.66&9.13&0.65&0.06&10.35&2.66&2.41&0.81&2.21&6.07\\
U7901&10.22&3.16&9.79&1.84&0.27&10.36&3.18&1.46&0.63&4.28&9.06\\
U8279&10.44&2.87&10.09&1.65&0.31&10.60&2.90&2.67&0.80&2.65&5.95\\
U8865&10.63&2.26&9.77&0.58&0.12&10.69&2.27&5.65&0.50&6.03&11.84\\
U9481&10.65&2.39&10.57&1.45&0.45&10.91&2.44&6.49&0.85&2.41&6.65\\
U9926&10.76&3.05&9.86&1.33&0.11&10.81&3.06&2.61&0.76&2.51&5.15\\
U9943&10.65&2.92&9.99&1.42&0.18&10.73&2.94&2.86&0.80&2.48&4.87\\
U10083&10.34&2.59&9.79&1.20&0.22&10.45&2.60&3.02&0.83&2.29&3.51\\
U10445&9.59&2.36&9.56&1.40&0.48&9.87&2.41&2.13&0.70&3.88&8.99\\
U11628&11.43&2.82&10.38&0.99&0.08&11.47&2.83&6.92&0.67&2.99&9.83\\
U11872&10.48&3.40&9.89&2.05&0.20&10.58&3.42&1.28&0.89&1.73&4.99\\
U12343&11.19&2.64&10.53&1.11&0.18&11.28&2.65&7.52&0.69&3.44&9.79\\
U12511&10.30&2.28&10.63&1.83&0.68&10.80&2.41&5.89&0.58&5.15&19.15\\
U12614&10.93&2.71&10.32&1.15&0.20&11.03&2.72&5.60&0.84&2.71&5.17\\
U12638&10.93&2.28&10.42&0.88&0.24&11.05&2.30&8.98&0.58&5.17&13.91\\
U12654&10.57&2.45&10.34&1.31&0.37&10.77&2.48&5.43&0.74&3.37&7.42\\
U6399&9.11&1.78&9.17&0.88&0.54&9.44&1.83&2.57&0.49&7.99&21.95\\
U6446&9.12&2.22&9.42&1.32&0.66&9.60&2.27&2.05&0.72&4.58&15.14\\
N3726&10.26&2.60&10.22&1.57&0.48&10.54&2.64&3.63&0.77&3.35&5.51\\
N3769&9.81&2.73&9.94&1.96&0.57&10.18&2.80&1.91&0.88&2.40&8.36\\
N3877&10.52&2.92&9.60&1.09&0.11&10.57&2.93&2.55&0.90&2.24&4.12\\
N3893&10.47&2.89&10.15&1.70&0.32&10.64&2.92&2.71&0.80&2.64&5.32\\
N3917&10.09&2.36&9.57&0.91&0.23&10.20&2.37&3.21&0.64&4.47&7.42\\
N3949&10.07&2.89&9.79&1.78&0.35&10.25&2.92&1.70&0.76&2.86&4.40\\
N3953&10.98&2.96&9.90&0.99&0.08&11.01&2.96&3.95&0.87&2.23&5.83\\
N3972&9.81&2.62&9.43&1.26&0.30&9.96&2.63&1.88&0.70&4.14&8.39\\
U6917&9.64&2.01&9.40&0.83&0.37&9.84&2.04&3.13&0.53&6.65&13.62\\
U6923&9.44&2.21&9.07&0.92&0.30&9.59&2.23&1.85&0.72&3.51&6.50\\
N3992&10.82&2.80&10.25&1.27&0.21&10.93&2.81&4.56&0.64&4.72&8.98\\
U6983&9.73&2.06&9.82&1.22&0.55&10.08&2.12&3.78&0.65&4.46&12.68\\
N4051&10.22&2.79&9.75&1.35&0.26&10.34&2.81&2.38&0.82&2.98&4.47\\
N4088&10.49&2.85&10.12&1.56&0.30&10.64&2.87&2.99&0.84&2.59&5.32\\
\hline\noalign{\smallskip}
\end{tabular}
\end{table*}

\setcounter{table}{1}

\begin{table*}
  \caption[]{Continued.}

\begin{tabular}{cccccccccccc} 
\hline\noalign{\smallskip}

Name  &Lg\ms$^{\mathrm{a}}$ & Lg\sigs$^{\mathrm{b}}$ & 
Lg\mg$^{\mathrm{c}}$ & Lg$\Sigma_{g}^{\mathrm{d}}$ & f$_{g}^{\mathrm{e}}$ & 
Lg\md$^{\mathrm{f}}$ & Lg\sigd$^{\mathrm{g}}$ & \hd$^{\mathrm{h}}$ & 
\vdvt$^{\mathrm{i}}$ & \mdynmd$^{\mathrm{j}}$ &  \mdynl$^{\mathrm{k}}$ 
\\
\hline\noalign{\smallskip}
N4100&10.53&2.92&10.12&1.55&0.28&10.67&2.93&2.93&0.84&2.73&5.17\\
N4102&10.55&3.34&9.32&1.31&0.06&10.58&3.34&1.39&0.91&1.67&6.76\\
N3718&10.64&2.33&10.01&0.89&0.19&10.73&2.35&5.42&0.47&6.67&11.35\\
N3729&10.31&3.11&9.29&1.11&0.09&10.35&3.11&1.69&1.02&1.91&5.50\\
U6773&9.01&1.90&8.70&0.70&0.33&9.19&1.93&1.62&0.68&3.85&7.09\\
U6818&9.05&1.97&9.13&1.08&0.55&9.39&2.03&1.94&0.61&5.26&14.72\\
N3985&9.44&2.66&8.98&1.21&0.26&9.57&2.67&1.15&0.83&2.95&3.81\\
U7089&9.28&1.82&9.40&0.92&0.57&9.65&1.87&3.18&0.72&4.00&11.50\\
U7094&8.74&1.93&8.92&1.08&0.60&9.14&1.99&1.55&1.05&1.91&6.32\\
N4117&9.53&2.22&9.12&1.22&0.28&9.67&2.26&1.55&0.38&7.80&22.07\\
N4138&10.40&3.28&8.85&0.94&0.03&10.41&3.28&1.22&0.84&1.94&7.04\\
N4218&8.97&2.74&8.63&1.46&0.32&9.13&2.76&0.61&0.80&2.92&3.42\\
N4220&10.30&3.16&9.13&1.10&0.06&10.32&3.17&1.43&0.74&3.15&5.39\\
U128&10.05&1.62&10.16&0.73&0.56&10.41&1.67&9.57&0.41&12.24&35.99\\
F563-V2&9.54&2.11&9.60&1.18&0.54&9.87&2.16&2.93&0.87&2.66&7.34\\
F568-3&9.85&1.96&9.82&0.95&0.49&10.14&2.00&4.75&0.59&5.87&14.11\\
F574-1&9.86&1.65&9.83&0.82&0.49&10.15&1.71&6.07&0.58&4.87&12.36\\
F583-1&9.11&1.67&9.52&1.43&0.72&9.67&1.87&2.94&0.55&5.56&26.16\\
MilkyWay&10.64&2.82&10.09&1.38&0.22&10.75&2.83&3.40&0.71&3.42&10.50\\
Andromeda&11.09&3.07&10.05&1.08&0.08&11.13&3.07&4.26&0.87&2.54&6.83\\
\hline\noalign{\smallskip}
\end{tabular}

\begin{list}{}{}
\item[$^{\mathrm{a}}$] Total stellar mass (disk+bulge) in \msun. 
\item[$^{\mathrm{b}}$] Disk stellar central surface density in \msun pc$^{-2}$.
\item[$^{\mathrm{c}}$] Total gas mass in \msun.
\item[$^{\mathrm{d}}$] Disk gas central surface density in \msun pc$^{-2}$.
\item[$^{\mathrm{e}}$] Gas fraction (\mg/[\mg\ + \ms]).
\item[$^{\mathrm{f}}$] Total baryonic mass (\ms\ + \mg) in \msun.
\item[$^{\mathrm{g}}$] Disk total central surface density in \msun pc$^{-2}$.
\item[$^{\mathrm{h}}$] Disk (stellar + gaseous) scale radius in kpc.
\item[$^{\mathrm{i}}$] Disk-to-total maximum velocity ratio.
\item[$^{\mathrm{j}}$] Dynamical-to-baryonic mass ratio at 5\hd. 
\item[$^{\mathrm{k}}$] Dynamical mass-to-light ratio in the band $B$ at 5\hd
(\msun / $L_{B\msun}$. 

\end{list}
\end{table*}

\end{document}